\documentclass[aps,prx,reprint,groupedaddress,amsmath,amssymb]{revtex4-2}

\usepackage[]{hyperref}
\usepackage[]{braket}
\usepackage{graphicx}
\usepackage{enumitem}
\usepackage{xr}

\newcommand{\widebar}[1]{\bar{#1}}

\newcommand{\dd}{{\mathrm{d}}}

\newcommand{\Ncomp}{{N_\mathrm{c}}}
\newcommand{\Nphase}{{N_\mathrm{p}}}
\newcommand{\chimean}{{\widebar{\chi}}}
\newcommand{\chistd}{{\sigma_{\chi}}}
\newcommand{\barphi}{{\bar{\phi}}}
\newcommand{\barNphase}{{\widebar{N}_\mathrm{p}}}
\newcommand{\Nunstable}{{N_\mathrm{u}}}
\newcommand{\barNunstable}{{\widebar{N}_\mathrm{u}}}

\newcommand{\RN}[1]{%
  \textup{\uppercase\expandafter{\romannumeral#1}}%
}

\newcommand{\TODO}[1]{}

\newcommand{\Eqref}[1]{Eq.~(\ref{#1})}  
\newcommand{\Eqrefst}[1]{Eq.~(\ref*{#1})}
\newcommand{\Seqref}[1]{Seq.~\eqref{#1}}
\newcommand{\Eqsref}[1]{Eqs.~\eqref{#1}}
\newcommand{\Figref}[1]{Fig.~\ref{#1}}
\newcommand{\Figrefst}[1]{Fig.~\ref*{#1}}

\newcommand{\SFigref}[1]{Supporting Fig.~#1}

\newcommand{\supportingName}{Appendix}
\newcommand{\paperTitle}{Generic multicomponent mixtures are multistable}

\hypersetup{
  colorlinks   = true,
  urlcolor     = blue,
  linkcolor    = blue,
  citecolor   = red
}

\begin{document}

\title{\paperTitle{}}

\author{Yicheng Qiang}
\affiliation{Max Planck Institute for Dynamics and Self-Organization, Am Fa{\ss}berg 17, 37077 G{\"o}ttingen, Germany}
\author{Chengjie Luo}
\affiliation{Max Planck Institute for Dynamics and Self-Organization, Am Fa{\ss}berg 17, 37077 G{\"o}ttingen, Germany}
\author{David Zwicker}
\email[]{david.zwicker@ds.mpg.de}
\affiliation{Max Planck Institute for Dynamics and Self-Organization, Am Fa{\ss}berg 17, 37077 G{\"o}ttingen, Germany}

\begin{abstract}
Liquid mixtures of many interacting components often exhibit numerous coexisting types of droplets. An exciting example is the cytosol of biological cells, where diverse droplets, called condensates, are essential for cellular function. However, how much their formation is constrained by thermodynamics is currently unclear. Linear stability analysis predicts that homogeneous mixtures become more robust to fluctuations as the number of components increases, suggesting that droplets do not form easily in multicomponent mixtures. In contrast, we show through numerical simulations and analytical scaling laws that the number of coexisting phases typically increases with the number of components in equilibrium. The combination of both results suggests that generic multicomponent mixtures can maintain many metastable states with various droplets, generalizing the nucleation-and-growth regime of binary mixtures. Our theory also indicates why cells exhibit much fewer condensates than components and how they could exploit multistability to independently form various condensates.
\end{abstract}

\maketitle

Mixtures with many components are ubiquitous in nature.
One particularly interesting case is the cytosol of biological cells, which contains millions of different molecules~\cite{cooper2022Cell}.
Despite this complexity, cells fulfill their function flexibly and robustly.
In part, this is because they leverage the physics of phase separation to form various biomolecular condensates~\cite{hyman2014LiquidLiquid,banani2017Biomolecular,jacobs2017Phase,sartori2020Lessons,lafontaine2021Nucleolus,azaldegui2021Emergence}.
However, it is not clear how cells control which condensates form.
In particular, it is unclear how much can be attributed to equilibrium thermodynamics and how many condensates are maintained by active processes.
Theoretical arguments based on linear stability analysis show that homogeneous states become more robust to fluctuations with larger component counts~\cite{sear2003Instabilities,jacobs2013Predicting,shrinivas2021Phase,thewes2023Composition}, suggesting that the cytosol should stay homogeneous~\cite{sear2005Cytoplasm}.
However, such stability analysis cannot assess consequences of large fluctuations, which could lead to the formation of various droplets.
Such behavior is familiar from simple binary mixtures in the nucleation--and--growth regime, which explains many aspects of the formation of individual in cells~\cite{hyman2014LiquidLiquid,berry2018Physical}. %
Yet, it is unclear whether such behavior is typical for mixtures of many components.

We here study the generic behavior of multicomponent mixtures and investigate whether there is a generalized nucleation--and--growth regime.
While the stability of homogeneous states has been discussed before~\cite{sear2003Instabilities,jacobs2013Predicting,shrinivas2021Phase,thewes2023Composition}, phase-separated states have been analyzed less thoroughly.
This is mainly because computing high-dimensional phase diagrams is challenging~\cite{mao2019Phase,mao2020Designing,jacobs2023Theory}.
To overcome this limitation, we combine a numerical method detecting phase-separated equilibrium states with analytically derived scaling laws for the number of coexisting phases. %
Our analysis shows that the number of coexisting phases in equilibrium typically increases with component count, while the homogeneous state remains metastable.
This result applies for a large range of interaction strengths between components and for various overall compositions of the system.
Taken together, this implies that the corresponding nucleation--and--growth region expands with component count. %

\subsection*{Generic equilibrium model of multicomponent mixtures}%

To analyze generic multicomponent mixtures, we consider an incompressible, isothermal system that is occupied by $\Nphase$ different homogeneous phases.
Each phase~$\alpha$ is characterized by a relative volume $J_\alpha$ and the volume fractions $\phi_i^{(\alpha)}$ of all components $i$ in this phase.
Together, all phases fill the entire system, implying the constraint $\sum_{\alpha=1}^{\Nphase} J_\alpha=1$, and mass conservation additionally fixes the average volume fractions $\barphi_i = \sum_{\alpha=1}^{\Nphase} J_\alpha \phi_i^{(\alpha)}$.
The system's behavior is governed by the average free energy
\begin{equation}
    \bar{f}(\Nphase, \{J_\alpha\}, \{\phi_i^{(\alpha)}\}) = \sum_{\alpha=1}^{\Nphase} J_\alpha f(\{\phi_i^{(\alpha)}\})
    \label{M-eqn:fe_many}
    \;,
\end{equation}
where  we neglect interfacial contributions assuming a thermodynamically large system.
To determine equilibrium states, which correspond to global minima of $\bar f$, we next introduce the free energy density~$f$ of a mixture of $\Ncomp$ components.
For simplicity we consider a non-dimensionalized Flory-Huggins free energy with equal molecular volumes~\cite{flory1942Thermodynamics,huggins1941Solutions},
\begin{equation}
    f(\{\phi_i\}_{i=2}^\Ncomp) = \sum_{i=1}^\Ncomp \phi_i \ln \phi_i + \frac{1}{2}\sum_{i,j=1}^\Ncomp \chi_{ij} \phi_i \phi_j
    \label{M-eqn:fe_single_general}
    \;,
\end{equation}
where $\phi_1 = 1-\sum_{i=2}^\Ncomp\phi_i$ since the mixture fills space. %
Here, the first term captures the translational entropy of all components and the second term accounts for interactions via the symmetric Flory matrix $\chi_{ij} = \chi_{ji}$ with $\chi_{ii}=0$.
Its entries $\chi_{ij}$ quantify the interaction energy between species $i$ and $j$ in units of the thermal energy $k_\mathrm{B}T$ per unit volume.

Given an interaction matrix~$\chi_{ij}$ and average composition $\barphi_i$, we can in principle minimize $\bar f$ to obtain the phase count $\Nphase$ together with the fractions of volume $J_\alpha$ and compositions $\phi_i^{(\alpha)}$ of all phases.
However, this task is complex since generic multicomponent mixtures vary in the number of components ($\Ncomp$), their average composition ($\barphi_i$), and the pairwise interactions ($\chi_{ij}$).
To increase complexity step by step, we first study a very symmetric system and then relax the associated constraint on $\barphi_i$ and $\chi_{ij}$ successively.

\subsection*{Symmetric mixtures exhibit a discrepancy between equilibrium and stability of homogeneous states}
To build intuition for the behavior of multicomponent mixtures, we start by analyzing the special case of identical interactions ($\chi_{i\neq j} = \chimean$) and symmetric compositions ($\barphi_i = 1/\Ncomp$), so we only vary $\chimean$ and the component count $\Ncomp$.

We first analyze homogeneous states and ask when they become unstable.
This is the case when the Hessian matrix $\boldsymbol{H} = \{H_{ij}\}_{i,j=2}^{\Ncomp}$ with $H_{ij}= \partial^2 f/\partial \barphi_i \partial \barphi_j$ is no longer positive definite, i.e., when at least one eigenvalue is negative.
For $\chi_{i\neq j} = \chimean$ and $\barphi_i = 1/\Ncomp$, the Hessian matrix simplifies to $H_{ij}=(\delta_{ij}+1) (\Ncomp - \chimean )$.
The associated eigenvalues, $\Ncomp - \chimean$ and $\Ncomp(\Ncomp - \chimean)$, are negative when
\begin{equation}
    \chimean>\Ncomp
    \;.
    \label{M-eqn:spinodal_symchi_symphi}
\end{equation}
If this condition is fulfilled, the homogeneous state is unstable and phase separation is inevitable.
Conversely, the homogeneous state is robust to small fluctuations for $\chimean < \Ncomp$.
However, large fluctuations could drive the system to a completely different state.
In particular, phase separation could still happen if there exists another state with multiple phases that has lower free energy than the homogeneous state~\cite{mcmaster1973Aspects,zwicker2022Intertwined}.
In this case, the homogeneous state would only be metastable, even if $\chimean < \Ncomp$.

To determine when phase separation happens, we next discuss equilibrium states, which correspond to global minima of the free energy density $\bar f$ given by \Eqref{M-eqn:fe_many}.
In the symmetric case considered in this section, the only possible phase-separated state in equilibrium is the one where $\Ncomp$ phases coexist, each concentrating one component to a high fraction $\phi^+$ while diluting the others to $\phi^- = (1 - \phi^+)/(\Ncomp - 1)$.
Consequently, the minimal interaction strength to have phase separation in equilibrium can be determined numerically by comparing the free energy of this phase-separated state with the homogeneous state.
To get an estimate for the minimal interaction strength, we use a variational principle: we identify a suboptimal phase-separated state and require it to have lower energy than the homogeneous state. %
Since the state where each of the $\Ncomp$ phases enriches exactly one component ($\phi^+=1, \phi^-=0$) has $\bar f=0$, the equilibrium state must be a phase-separated state when the homogeneous state has $\bar f>0$, implying the sufficient condition
\begin{equation}
    \chimean > \frac{2\Ncomp}{\Ncomp-1}\ln \Ncomp
    \;.
    \label{M-eqn:binodal_symchi_symphi_approx}
\end{equation}
\SFigref{M-1} shows that \Eqref{M-eqn:binodal_symchi_symphi_approx} provides a good approximation for the numerically determined critical value of $\chimean$, especially for large $\Ncomp$.
A more detailed analysis presented in the \supportingName{} shows that the diluted components have the fraction
\begin{equation}
    \phi^-(\Ncomp)  \approx - \frac{W_0\!\left[-e^{-\chimean}\left(\Ncomp(\chimean -1) + 1\right)\right]}{\Ncomp(\chimean -1) + 1}
    \overset{\chimean \gg 1}{\approx} e^{-\chimean}\;,
    \label{M-eqn:phi_star_approx}
\end{equation}
where $W_0(r)$ is the principal branch of the Lambert $W$ function satisfying $W_0(r)e^{W_0(r)}=r$.
The limit for large $\chimean$ shows that dilute fractions are suppressed exponentially.

These results confirm the expectation that stronger repulsion (i.e., larger $\chimean$) destabilizes the homogeneous state (\Eqref{M-eqn:spinodal_symchi_symphi}), induces phase separation (\Eqref{M-eqn:binodal_symchi_symphi_approx}), and enhances the compositional differences between phases (\Eqref{M-eqn:phi_star_approx}).
However, comparing Eqs.~(\ref{M-eqn:spinodal_symchi_symphi}) and (\ref{M-eqn:binodal_symchi_symphi_approx}) reveals a huge discrepancy between the stability analysis and equilibrium:
While the typical interaction $\chimean$ needs to scale with the component count $\Ncomp$ to destabilize the homogeneous state, phase separation is already possible when $\chimean$ scales only logarithmically with~$\Ncomp$.
These scalings already suggest that the parameter region where the homogeneous state is metastable but the equilibrium state exhibits phase separation (corresponding to the nucleation--and--growth regime) expands with larger component count~$\Ncomp$.

\subsection*{Compositional variation maintains the discrepancy}
Phase separation of multicomponent mixtures not only depends on interactions but also on composition~\cite{thewes2023Composition}.
We thus next test how compositional variations affect the discrepancy between unstable homogeneous states and equilibrium.
To capture compositional variations in an unbiased way, we average over all permissible volume fractions ($\barphi_i \ge 0$, $\sum_{i=1}^\Ncomp \barphi_i=1$), maintaining identical interactions for now ($\chi_{i\neq j} = \chimean$).
Since this system is more variable than the one discussed above, we can distinguish various degrees of instability of the homogenous state by counting the number of negative eigenvalues of the Hessian~$\boldsymbol H$, denoted as the number of unstable modes, $\Nunstable$.
If the homogeneous state is the equilibrium state, we necessarily have $\Nunstable=0$ and $\Nphase=1$.
In contrast, for very strong phase separation, the homogeneous state will be maximally unstable, $\Nunstable=\Ncomp-1$, and the phase count assumes the maximal value $\Nphase=\Ncomp$ allowed by Gibbs phase rule~\cite{gibbs1876Equilibrium}.
In both extremes we have $\Nphase = \Nunstable + 1$, but these two values might deviate from each other substantially in intermediate parameter regions.
To quantify this behavior, we next study the mean values $\barNphase$ and $\barNunstable+1$ averaged over all permissible compositions.

\begin{figure}[t]
    \centering
    \includegraphics[width=\linewidth]{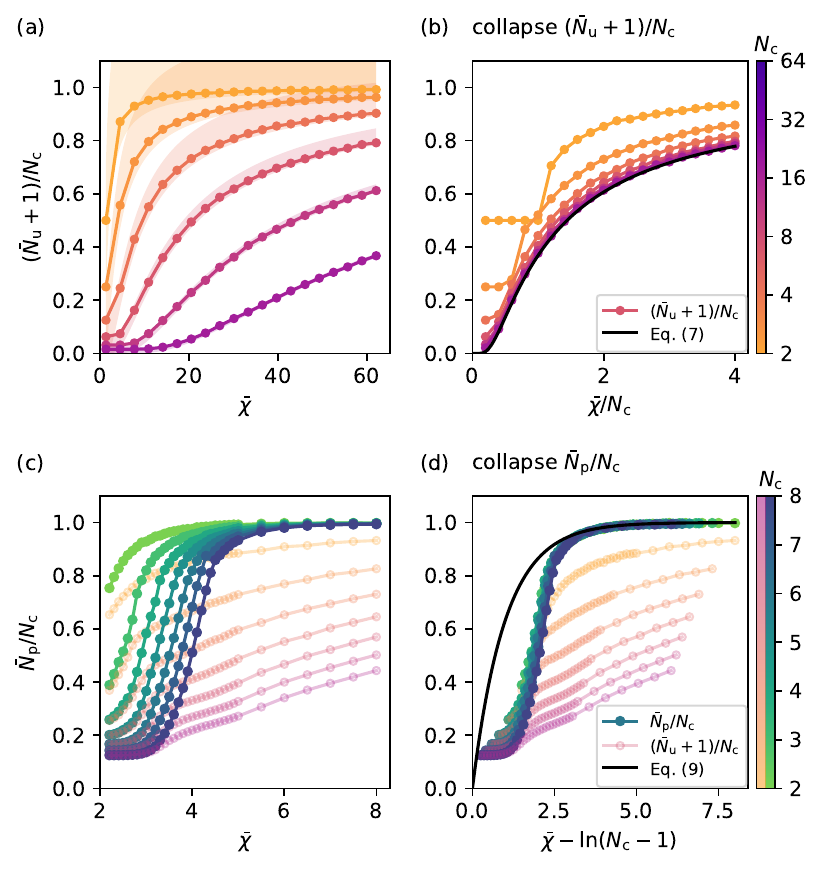}
    \caption{
        \textbf{Average number of phases exceeds number of unstable modes for identical interactions.}
        (a, b) Normalized average number of unstable modes, $(\barNunstable+1)/\Ncomp$, as a function of the mean interaction strength $\chimean$ in (a) and the scaled interaction $\chimean/\Ncomp$ in (b) for various component counts $\Ncomp$.
        The shaded area in (a) indicates the bounds given by \Eqref{M-eqn:unstable_modes_count_bounds}, and the black line in (b) marks the asymptotic expression (\ref{M-eqn:unstable_modes_count_asymptotic}).
        (c, d) Normalized average number of coexisting phases, $\barNphase/\Ncomp$, (green-blue) and $(\barNunstable+1)/\Ncomp$ (orange-purple) as a function of $\chimean$ in (c) and $\chimean - \ln(\Ncomp-1)$ in (d).
        The black line in (d) marks the asymptotic expression \eqref{M-eqn:phase_count_asymptotic}.
        (a--d) %
        Each dot results from an average over $500$--$20\,000$ uniformly sampled compositions $\barphi_i$.
    }
    \label{M-fig:combined_sym}
\end{figure}

\begin{figure*}[t]
    \centering
    \includegraphics[width=0.9\linewidth]{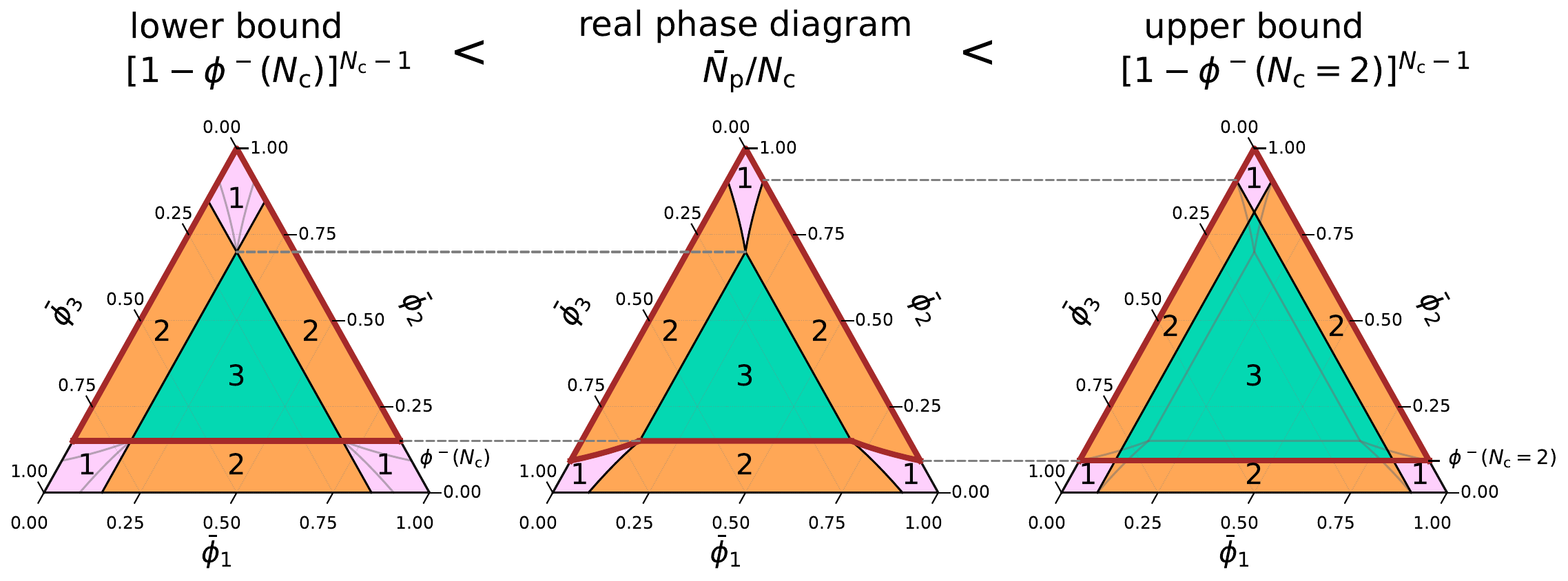}
    \caption{
        \textbf{Illustration of phase count bounds.}
        The central panel shows the phase diagram as a function of the volume fractions~$\phi_i$ of $\Ncomp=3$ components for $\chi_{i\neq j}=2.8$.
        The colored regions indicate the number of coexisting phases, $\Nphase=1,2,3$.
        The adjacent panels approximate the phase diagram using boundaries parallel to the axes, obtained by expanding the $\Nphase=1$ region (left) and the central region with $\Nphase=\Ncomp$ (right).
        They respectively provide lower and upper bounds to the average phase count $\barNphase$, which can be obtained from the fraction covered by an area associated with each component (thick red lines).
        The average phase count~$\barNphase$ for the three panels is $2.17$, $2.22$, and $2.47$ (from left to right).
    }
    \label{M-fig:binodal_sym}
\end{figure*}

\subsubsection*{Average number of unstable modes}
To determine the average number of unstable modes, $\barNunstable$, we diagonalized the Hessian matrix $\boldsymbol{H}$ numerically for many different compositions, sampled uniformly over the entire phase diagram~\cite{zwicker2022Evolved}.
\Figref{M-fig:combined_sym}a shows that $\barNunstable$ increases with the mean interaction strength~$\chimean$, as expected.
An analytical estimate for $\barNunstable$ follows from Cauchy's interlacing theorem~\cite{bellman1997Introduction}:
The number $\Nunstable$ of negative eigenvalues of $\boldsymbol{H}$ equals the number of sign changes in the sequence
$\{\det{\boldsymbol{H}_0}, \det{\boldsymbol{H}_1},\hdots, \det{\boldsymbol{H}_{\Ncomp -1}}\}$,
where $\det{\boldsymbol{H}_n}$ is the determinant of the rank-$n$ leading minor of $\boldsymbol{H}$.
For identical interactions ($\chi_{i\neq j} = \chimean$), we obtain $\det{\boldsymbol{H}_n} = (\prod_{i=1}^{n+1} q_i)(\sum _{i=1}^{n+1} q_i^{-1})$ with $q_i = \barphi_i^{-1} -\chimean$.
The number of sign changes in the sequence is thus either $N_-$ or $N_--1$, where $N_-$ is the number of negative $q_i$, since each negative $q_i$ contributes one sign change in the product $\prod_{i=1}^{n+1} q_i$ when varying $n$, while a potential sign change of the second factor, $\sum _{i=1}^{n+1} q_i^{-1}$, cancels at most one sign change; see \supportingName{} for a detailed derivation.
Since $N_-$ simply counts how many $\barphi_i$ are larger than $\chimean^{-1}$, $\barNunstable$ is bounded by  $E_{\Ncomp}[\barphi_i>\chimean^{-1}] \leq \barNunstable + 1 \leq E_{\Ncomp}[\barphi_i>\chimean^{-1}]+1$, where $E_{\Ncomp}[\barphi_i>h] = \Ncomp (1-h)^{\Ncomp-1}$ denotes the expected number of $\barphi_i$ larger than $h$ when $\barphi_i$ is drawn randomly on the phase diagram.
Taken together, the normalized average number of unstable modes, $(\barNunstable+1)/\Ncomp$, is  bounded by
\begin{equation}
    \left(1-\frac{1}{\chimean}\right)^{\Ncomp-1}
    \leq \frac{\barNunstable +1}{\Ncomp}
    \leq \left(1-\frac{1}{\chimean}\right)^{\Ncomp-1} + \frac{1}{\Ncomp}
    \;,
    \label{M-eqn:unstable_modes_count_bounds}
\end{equation}
consistent with numerical data (\Figref{M-fig:combined_sym}a).
The two bounds in \Eqref{M-eqn:unstable_modes_count_bounds} converge for large $\Ncomp$, implying
\begin{equation}
    \frac{\barNunstable +1}{\Ncomp}
    \overset{\Ncomp\gg1}{\approx} \left(1-\frac{1}{\chimean}\right)^{\Ncomp-1}
    \overset{\chimean\gg1}{\approx} e^{-\Ncomp/\chimean}
    \;.
    \label{M-eqn:unstable_modes_count_asymptotic}
\end{equation}
The limit on the right suggests that the normalized number of unstable modes averaged over the phase diagram, $(\barNunstable + 1)/\Ncomp$, only depends on the single scaled interaction parameter $\chimean/\Ncomp$. %
The data collapse shown in \Figref{M-fig:combined_sym}b indeed confirms this scaling for $\Ncomp \gtrsim 16$.

\subsubsection*{Average number of coexisting phases}

To test how well the average number~$\barNunstable$ of unstable modes predicts the average phase count, $\barNphase$, we next investigate equilibrium states by numerically minimizing the  mean free energy $\bar f$ given by \Eqref{M-eqn:fe_many}.
We designed an efficient algorithm to determine coexisting phases for arbitrary interactions and mean compositions~\cite{qiang2025Flory}, which is similar to the Gibbs ensemble method~\cite{mester2013Numerical,schmid1998Selfconsistentfield,panagiotopoulos1995Gibbs,smit1989Calculation}.
The method represents the system's state by many candidate phases called compartments.
It then minimizes $\bar{f}$ by redistributing components and volumes between compartments while obeying volume and material conservation. %
To ensure that we detect the global minimum of $\bar f$, we use many more compartments than the maximal phase count $N_\mathrm{p}^\mathrm{max}=\Ncomp$ imposed by Gibbs phase law.
Consequently, many compartments will attain very similar compositions at equilibrium, and we combine them to detect the number $\Nphase$ of distinct phases; see \supportingName{} for  numerical details.
In a key improvement over our earlier implementation~\cite{zwicker2022Evolved,jacobs2023Theory}, we now initialize compartments with controlled average fractions~$\barphi_i$.
This allows us to sample phase counts~$\Nphase$ uniformly over the entire phase diagram %
to estimate the average number of coexisting phases, $\barNphase$, in equilibrium.

The numerical results shown in \Figref{M-fig:combined_sym}c indicate that~$\barNphase$ increases much more quickly as a function of the interaction strength~$\chimean$ than the mode count~$\barNunstable$.
Moreover, the scaled interaction strength $\chimean/\Ncomp$, derived for $\barNunstable$ from \Eqref{M-eqn:unstable_modes_count_asymptotic}, cannot collapse these data.
To derive an alternative scaling law, we focus on strong interactions, where \Eqref{M-eqn:binodal_symchi_symphi_approx} is satisfied, so we have the maximal number of phases ($\Nphase=\Ncomp$) for equal composition at the center of the phase diagram ($\barphi_i=1/\Ncomp$).
The region where the $\Ncomp$-phase state is stable forms a $(\Ncomp-1)$-dimensional simplex (green region in \Figref{M-fig:binodal_sym}), which covers a larger area for higher $\chimean$. %
While fewer phases are observed outside this region, the symmetric structure of the phase diagram still allows us to determine the average phase count $\barNphase$.
In essence, the phase diagram has a three-fold rotational symmetry in the example of \Figref{M-fig:binodal_sym}, so we can focus on one primitive region (highlighted by the red outlines).
Since the phase count at each point in the phase diagram equals the number of (symmetric) primitive regions covering that point, $\barNphase/\Ncomp$ is simply given by the fraction of the phase diagram covered by a single primitive region.
We next derive lower and upper bounds by approximating the primitive region by inscribed and circumscribed simplexes.
For the example shown in the central panel of \Figref{M-fig:binodal_sym}, we thus replace the lower boundary of the primitive region by a straight line coinciding with the boundary of the central three-phase region (left panel in \Figref{M-fig:binodal_sym}).
The composition of the three-phase region is $\phi^-(\Ncomp)$ given by \Eqref{M-eqn:phi_star_approx}, so the relative area of the simplex is given by $[1-\phi^-(\Ncomp)]^{\Ncomp-1}$.
In contrast, for the upper bound we replace the boundary by a straight line anchored at the corner points of the primitive region (right panel in \Figref{M-fig:binodal_sym}).
The corresponding fraction is $\phi^-(\Ncomp=2)$ since the edge of the phase diagram corresponds to a binary system.
These arguments generalize to higher dimensions, so the normalized average phase count $\barNphase/\Ncomp$ is generally bounded by
\begin{equation}
    \left[1-\phi^-(\Ncomp)\right]^{\Ncomp-1}
    \le \frac{\barNphase}{\Ncomp}
    \le \left[1-\phi^-(\Ncomp=2)\right]^{\Ncomp-1}
    \;.
    \label{M-eqn:phase_count_bounds}
\end{equation}
The fraction $\phi^-(\Ncomp)$ given by \Eqref{M-eqn:phi_star_approx} converges to $e^{-\chimean}$ for all $\Ncomp$ if $\chimean \gg 1$, so
\begin{equation}
    \frac{\barNphase}{\Ncomp} \overset{\chimean\gg1}{\approx} \left(1-e^{-\chimean}\right)^{\Ncomp-1}
    \approx 1 -(\Ncomp-1)e^{-\chimean}
    \;.
    \label{M-eqn:phase_count_asymptotic}
\end{equation}
This approximation reveals that $\barNphase$ is asymptotically given by $E_{\Ncomp}[\barphi_i>e^{-\chimean}]$, i.e., the expected number of $\barphi_i$ larger than $e^{-\chimean}$.
This result is similar to that for $\barNunstable$ following from \Eqref{M-eqn:unstable_modes_count_asymptotic}, except the threshold is now $1/e^{\chimean}$ instead of $1/\chimean$, implying an exponential scaling when switching from stability analysis to equilibrium.

Using \Eqref{M-eqn:phase_count_asymptotic}, we predict that $\barNphase/\Ncomp$ is governed by the single shifted interaction parameter $\chimean-\ln(\Ncomp-1)$. %
Indeed, \Figref{M-fig:combined_sym}d shows that this scaling collapses the data in a surprisingly large parameter range, even when the limits used to derive  \Eqref{M-eqn:phase_count_asymptotic} are violated.
The data collapse that we identify in \Figref{M-fig:combined_sym}d indicates that the interaction parameter $\chi_{i\neq j}=\chimean$ needs to exceed roughly $3 + \ln(\Ncomp-1)$ to have many phases ($\barNphase \approx \Ncomp$).
In contrast, the stability analysis predicted a linear scaling of $\chimean$ with $\Ncomp$, so that the discrepancy between the phase count and the number of unstable modes increases with larger component count~$\Ncomp$.

\subsection*{Random interactions maintain the discrepancy}
Beside compositional variations realistic mixtures exhibit large variations in the strength of the pairwise interaction between components. %
To test whether variable interactions affect the discrepancy between unstable homogeneous states and equilibrium, we next consider random interaction matrices~$\chi_{ij}$.
To focus on general features, we draw the entries $\chi_{ij}$ for $i < j$ independently from the normal distribution $\mathcal{N}(\chimean, \chistd)$, which allows us to vary the typical interaction strength~$\chimean$ and the associated standard deviation~$\chistd$. %
In particular, this choice allows direct comparison with previous results on the stability of homogeneous states~\cite{sear2003Instabilities,jacobs2013Predicting,shrinivas2021Phase,thewes2023Composition}.

\subsubsection*{Vanishing mean interaction}

\begin{figure}[t]
    \centering
    \includegraphics[width=\linewidth]{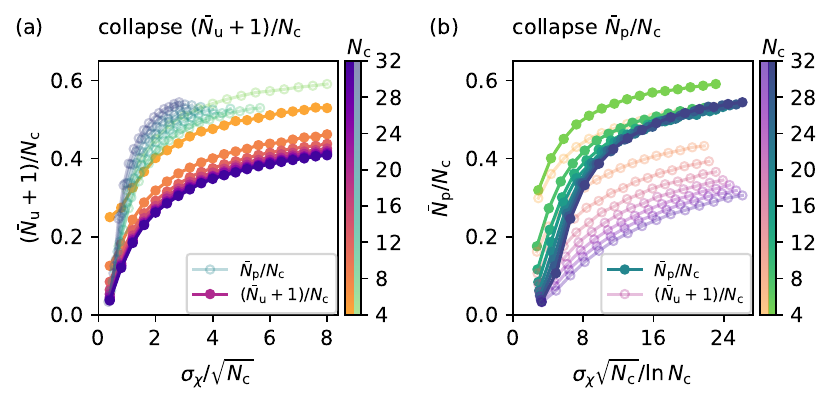}
    \caption{
        \textbf{More variable interactions elicit more phases.} %
        Normalized average number of unstable modes, $(\barNunstable+1)/\Ncomp$ (orange-purple), and normalized average number of coexisting phases, $\barNphase/\Ncomp$ (green-blue), as a function of two different scalings of the standard deviation $\sigma_\chi$ of the interactions for various component counts $\Ncomp$ and $\chimean=0$.
        Each dot results from an average over $10^3$--$10^4$ randomly chosen pairs of interaction matrices $\chi_{ij}$ and compositions $\barphi_i$.
    }
    \label{M-fig:binodal_random}
\end{figure}

For simplicity, we first vary only the standard deviation $\chistd$ of the normal distribution governing the symmetric interaction matrix~$\chi_{ij}$ while keeping a vanishing mean interaction ($\chimean=0$).
For symmetric compositions ($\barphi_i = 1/\Ncomp$), $\barNunstable$ can be obtained by treating the Hessian~$\boldsymbol{H}$ as a random matrix.
In particular, $\boldsymbol{H}$ consists of an enthalpic part given by the random $\chi_{ij}$ matrix and an entropic part given by a diagonal matrix with $\barphi_i^{-1}$ on the diagonal.
The corresponding eigenvalues read $\lambda_i + \barphi_i^{-1}$, where the distribution of eigenvalues $\lambda_j$ of $\chi_{ij}$ forms a semicircle of radius of $2 \sigma_{\chi}\sqrt{\Ncomp}$ centered at the origin~\cite{sear2003Instabilities}.
In particular, the
average number of negative eigenvalues is then  $\barNunstable=E_{\Ncomp}[\lambda_j > \barphi_i^{-1}]$. %
Geometrically, the fraction of the average number of unstable modes, $\barNunstable / \Ncomp$, %
thus equals the fraction of the semicircle right of $\barphi_i^{-1}$, which is a function of the single parameter $\sigma_{\chi}\sqrt{\Ncomp} / \barphi_i^{-1} = \sigma_{\chi} \Ncomp^{-1/2}$.
This scaling even persists for random compositions $\barphi_i$~\cite{thewes2023Composition}, where the above analysis does not hold.
\Figref{M-fig:binodal_random}a confirms that this scaling collapses data for $\barNunstable$, whereas it fails for the phase counts~$\barNphase$ that we determined using our numerical algorithm.

To also obtain a scaling relation for the phase count~$\barNphase$, we exploit an analogy between $\barNunstable$ and $\barNphase$ that we derived above for identical interactions ($\chi_{i \neq j} = \chimean$):
We found  $\barNunstable \sim E_{\Ncomp}[\chimean > \barphi_i^{-1}]$ and $\barNphase \sim E_{\Ncomp}[e^{\chimean} > \barphi_i^{-1}]$, indicating that the two quantities only differ in the dependence on $\chimean$ in the argument.
For random interactions, we thus speculate that the scaling of $\barNphase$ can be inferred from the result for $\barNunstable$ by substituting $e^{\lambda_i}$ for the eigenvalues $\lambda_i$.
This procedure suggests that $\barNphase/\Ncomp$ can be expressed as a function of the scaling parameter $\sigma_{\chi}\sqrt{\Ncomp}/\ln{\Ncomp}$ for sufficiently large $\Ncomp$, which is confirmed by the data collapse shown in \Figref{M-fig:binodal_random}b.
Thus, larger variations~$\sigma_{\chi}$ of the interactions generally lead to more phases, and $\barNphase/\Ncomp$ even grows with $\Ncomp$ when $\sigma_{\chi}$ is fixed.
In contrast, the stability analysis suggests that $\sigma_{\chi}$ needs to grow with $\sqrt{\Ncomp}$ to have the same fraction of unstable modes, so $\barNunstable$ severely underestimates the phase count $\barNphase$, particularly for large $\Ncomp$.

\subsubsection*{Finite mean interaction}

\begin{figure}[t]
    \centering
    \includegraphics[width=\linewidth]{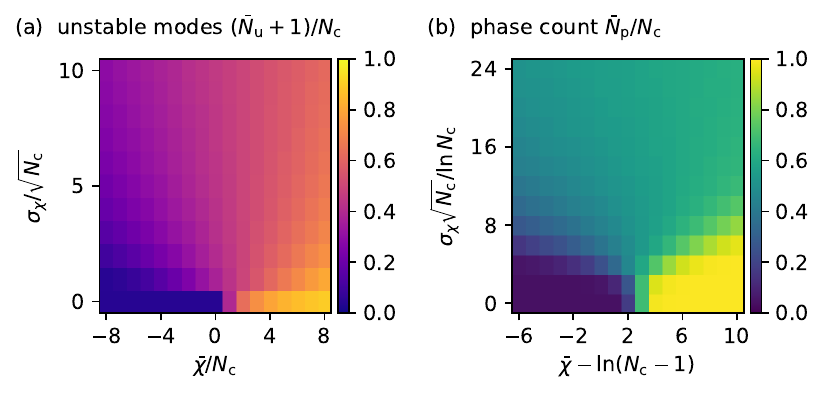}
    \caption{
        \textbf{Scaling functions for mixtures with random interactions.}
        (a) Normalized average number of unstable modes, $(\barNunstable+1)/\Ncomp$, as a function of scaled mean interaction~$\chimean$ and standard deviation $\chistd$.
        (b) Normalized average number of coexisting phases, $\barNphase/\Ncomp$, as a function of scaled $\chimean$ and $\chistd$.
        (a, b) The shown master functions were obtained by averaging $300$ samples of random interaction matrices and compositions for each value of $\chimean$ and $\chistd$ for component counts $\Ncomp=16,20,24,28,32$.
    }
    \label{M-fig:all_std_and_mean}
\end{figure}

We next study the impact of a non-zero mean interactions $\chimean$:
Combining the scalings for identical interactions ($\chistd=0$; see Eqs. \ref{M-eqn:unstable_modes_count_asymptotic} and \ref{M-eqn:phase_count_asymptotic}) and vanishing mean interactions ($\chimean=0$; see previous section), we postulate that the average number of unstable modes and phase count respectively scale as
\begin{subequations}
    \label{M-eqn:master_2D}
    \begin{align}
        \frac{\barNunstable + 1}{\Ncomp} & \approx
        g_\mathrm{u}\left(\frac{\chimean}{\Ncomp},\; \frac{\chistd}{\sqrt{\Ncomp}}\right) \label{M-eqn:master_2D_unstable}
        \qquad \text{and}
        \\
        \frac{\barNphase}{\Ncomp}        & \approx
        g_\mathrm{p}\left(\chimean  - \ln(\Ncomp-1),\; \frac{\chistd \sqrt{\Ncomp}}{\ln{\Ncomp}}\right), \label{M-eqn:master_2D_phase}
    \end{align}
\end{subequations}
where $g_\mathrm{u}$ and $g_\mathrm{p}$ are the associated master functions, %
which we determine numerically (\Figref{M-fig:all_std_and_mean}).
The deviation of the actually measured data to this prediction is generally low (see \SFigref{M-2}), %
implying that the proposed scaling captures the dominant behavior. %
The two master functions shown in \Figref{M-fig:all_std_and_mean} are qualitatively similar:
For strongly attractive interactions (large negative $\chimean$ and small $\chistd$), they are close to $0$, implying that only very few phases form.
In contrast, when repulsive interactions dominate (large $\chimean$ and small $\chistd$), the master functions converge to $1$, implying $\Ncomp$ phases coexist.
For strongly varying interactions (small $\chimean$ and large $\chistd$), both master functions converge to $\frac12$ and the respective influence of $\chimean$ can be approximated analytically; see \supportingName{}.
While the master functions exhibit similarities, the axes are scaled very differently, implying different interpretations of what constitutes large interactions:
For large $\Ncomp$, the mean interactions $\chimean$ and the variance $\sigma_\chi^2$ need to scale with $\Ncomp$ to obtain similar fractions $\barNunstable/\Ncomp$ of unstable modes.
In contrast, the predicted phase count $\barNphase$ exhibits a much weaker dependence on $\Ncomp$:
For instance, the influence of $\Ncomp$ on the mean interaction is only logarithmically.
Moreover, the predicted fraction $\barNphase/\Ncomp$ increases with $\Ncomp$ even if $\chistd$ is kept constant.
These scaling laws also hold for different samplings of the compositions (see \supportingName{}) and are thus likely robust.
In particular, we conclude that multicomponent mixtures typically have many more phases in equilibrium than the stability analysis of the homogeneous state predicts.

\subsection*{Multicomponent mixtures exhibit a large nucleation--and--growth regime}

\begin{figure}[t]
    \centering
    \includegraphics[width=\linewidth]{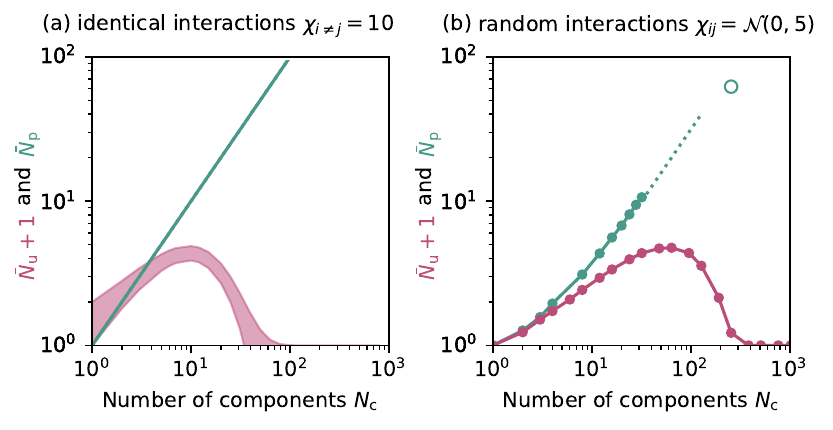}
    \caption{
        \textbf{Mixtures with metastable homogeneous states can have many coexisting phases in equilibrium.}
        (a) Analytical estimates of average number of coexisting phases $\barNphase$ (green line, \Eqref{M-eqn:phase_count_bounds}) and average number of unstable modes $\barNunstable$ (red region, \Eqref{M-eqn:unstable_modes_count_bounds}) as a function of component count $\Ncomp$ for mixtures with identical interactions ($\chi_{i \neq j} = 10$).
        (b) Numerically determined $\barNphase$ (green dots) and $\barNunstable$ (red dots) as a function of $\Ncomp$ for random interactions  ($\chimean = 0$, $\chistd=5$) and compositions.
        The green dotted line shows a linear extrapolation toward the single data point at $\Ncomp = 256$ (open circle).
    }
    \label{M-fig:counts_fixed_chi}
\end{figure}

The scaling laws given by \Eqref{M-eqn:master_2D} allow us to speculate about the expected number~$\Nunstable$ of unstable modes and the equilibrium phase count~$\Nphase$ in realistic systems.
As an example, we focus on biological cells, where a typical interaction strength is a few $k_\mathrm{B}T$, but the actual component count~$\Ncomp$ can vary over several orders of magnitude; there are an estimated number of $10^5$ different proteins~\cite{cooper2022Cell}.
For such a scenario, numerical data indicates that the homogeneous state becomes more robust ($\barNunstable$ approaches $0$)~\cite{sear2005Cytoplasm}, whereas the expected phase count in equilibrium increases linearly with $\Ncomp$ for identical interactions (\Figref{M-fig:counts_fixed_chi}a) and randomly distributed interactions (\Figref{M-fig:counts_fixed_chi}b).
This behavior can be understood based on the scaling laws given in \Eqref{M-eqn:master_2D}:
For large $\Ncomp$ at fixed $\chimean$ and $\chistd$, the fraction of unstable modes approaches $g_\mathrm{u}(0, 0) = 0$.
In contrast, the governing parameters for $\Nphase$ scale much slower, keeping the phase count fraction almost constant in a wide range of $\Ncomp$.
The number $\Nphase$ of coexisting phases would only decrease at much larger values of $\Ncomp$, implying a huge parameter range where the homogeneous state is metastable while the equilibrium state exhibits many phases. %

The discrepancy between the number of unstable modes and the equilibrium phase count generalizes the familiar nucleation--and--growth regime in binary mixtures.
Even in binary mixtures, the fraction beyond which phase separation sets in (the binodal) is approximately given by $e^{-\chi}$, whereas the homogeneous state only becomes unstable for fractions larger than $\frac12 \chi^{-1}$ (the spinodal)~\cite{zwicker2025Physics,qian2022Analytical}.
These scalings are consistent with the respective exponential and linear scalings that we revealed above, but we additional find a strong influence of the component count~$\Ncomp$.
Since $\Ncomp$ sets the dimensionality of the phase diagram, we speculate that its influence is of geometric origin:
The region where the homogeneous state is unstable typically occupies the central region of the phase diagram, so multistability might be more likely for parameters close to  boundaries.
The fraction of the phase diagram that is close to boundaries increases dramatically with dimension, suggesting that the behavior at boundaries dominates for large $\Ncomp$.

\subsection*{Multicomponent mixtures exhibit many metastable states}

\begin{figure}
    \centering
    \includegraphics[width=\linewidth]{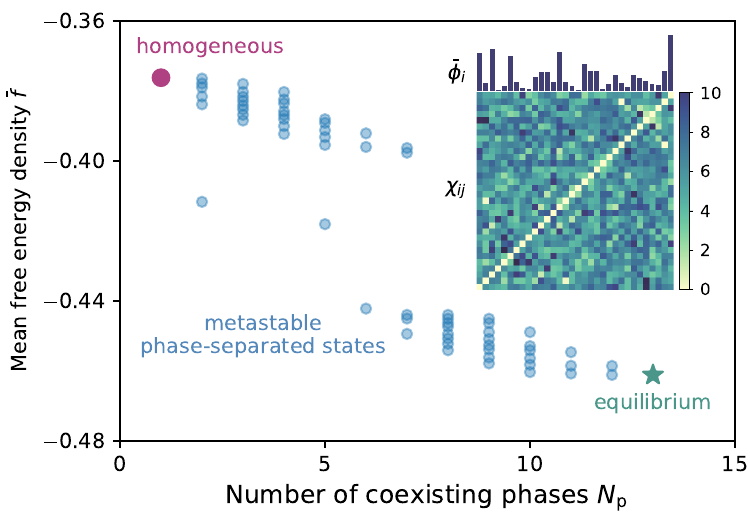}
    \caption{
        \textbf{Generic mixtures exhibit many metastable phase-separated states.}
        Mean free energy density~$\bar f$ for multiple (meta)stable states arranged by their phase count~$\Nphase$.
        The metastable homogeneous state (purple disk) has the highest free energy, whereas the free energy minimum corresponds to the equilibrium state with $13$ phases (green star).
        The inset shows the interaction matrix $\chi_{ij}$ and the average composition $\barphi_i$ of the investigated mixture with $\Ncomp=32$ components.
    }
    \label{M-fig:metastable_energies}
\end{figure}

We showed that many multicomponent mixtures have a metastable homogeneous state, while their equilibrium state exhibits many coexisting phases.
A system that starts in the homogeneous state thus needs to nucleate all these phases to reach equilibrium.
We speculate that this nucleation happens consecutively, implying the existence of intermediate metastable states with fewer coexisting phases than the equilibrium state.
To explore this exciting possibility, we use our numerical method~\cite{qiang2025Flory} to discover such intermediate states.
Briefly, we generate candidate states by either constraining the phase count or by removing phases from the equilibrium state.
In all cases, we then minimize $\bar f$ to find the nearest local free energy minimum, corresponding to a (meta)stable state.
Using this procedure, we find multiple metastable states between the homogeneous and the equilibrium state for a particular mixture with $\Ncomp=32$ components (\Figref{M-fig:metastable_energies}).
In this case, the metastable homogeneous state has the highest energy of all tested states.
The energy tends to decrease with the phase count~$\Nphase$, suggesting that phases can be nucleated successively to reach the equilibrium state with the lowest energy.
This analysis demonstrates that multicomponent mixtures typically exhibit many metastable states.

\subsection*{Surface tension does not impair multistability}

\begin{figure}
    \centering
    \includegraphics[width=\linewidth]{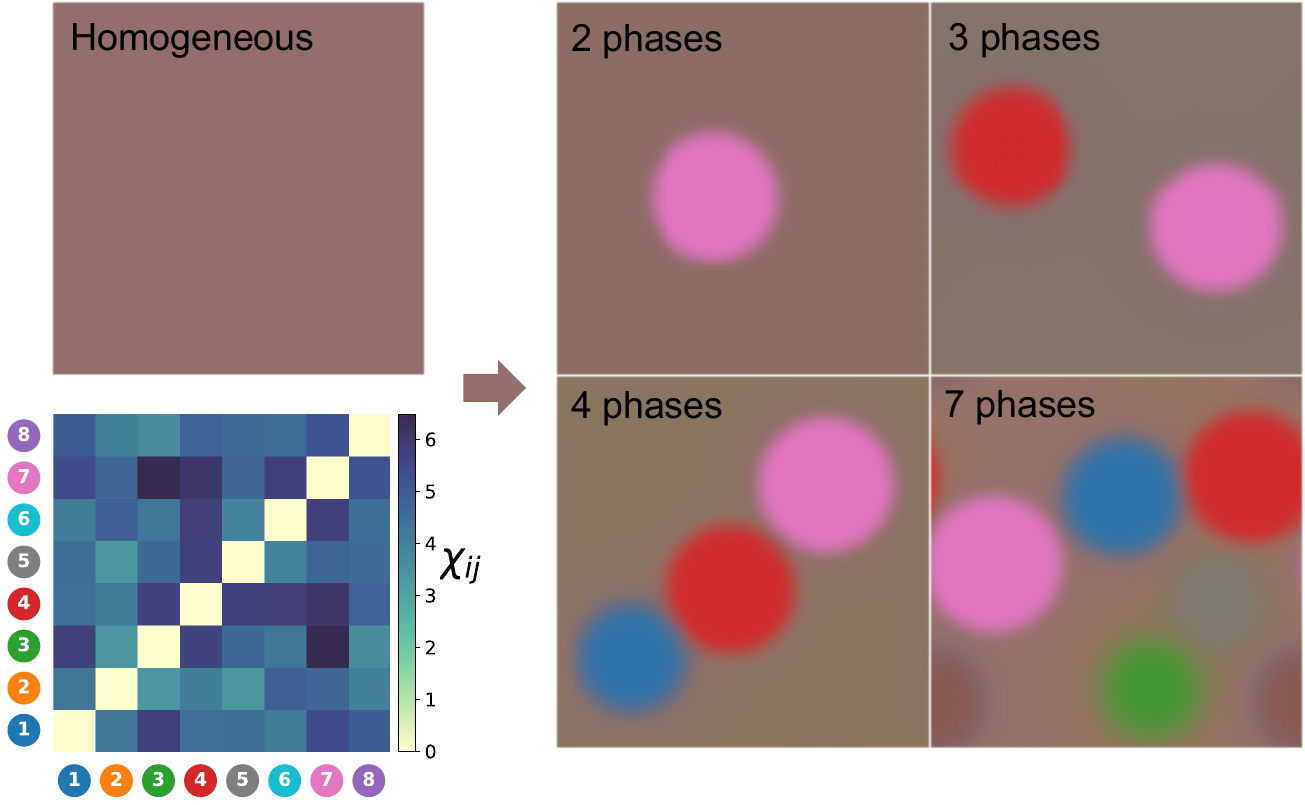}
    \caption{
        \textbf{Metastable states prevail in finite systems.}
        Density plots of (meta)stable configurations of a mixture with $\Ncomp=8$ components with interaction matrix $\chi_{ij}$ (lower left) and symmetric composition ($\barphi_i=\frac18$).
        Stationary configurations have been obtained by solving the Cahn-Hilliard equation; see \supportingName{}.
    }
    \label{M-fig:multistability}
\end{figure}
So far, we considered thermodynamically large systems, but biological cells are relatively small, so interfacial effects could matter. %
To check whether generic multicomponent mixtures still exhibit multistability in this case, we next consider spatially extended volume fraction fields $\phi_i(\boldsymbol r, t)$ in a periodic system conserving the average fractions $\barphi_i=\langle\phi_i(\boldsymbol r, t)\rangle_{\boldsymbol r}$.
The corresponding gradient descent dynamics that minimize the overall free energy are given by the Cahn-Hilliard equation~\cite{cahn1958Free,zwicker2025Physics}, $\partial_t \phi_i = \sum_j \nabla \cdot[\Lambda_{ij} \nabla (\partial f/\partial \phi_j - \ell^2\nabla^2\phi_j)]$ for $i=2,\ldots,\Ncomp$, where the mobility matrix~$\Lambda_{ij}$ sets time scales and $\ell$ controls the width of interfaces and is related to surface tension~\cite{mao2019Phase}.
Evolving these coupled equations numerically from various initial configurations, we find stationary configurations (\Figref{M-fig:multistability}).
Similar to the example shown in \Figref{M-fig:metastable_energies}, the spatially extended system exhibits a metastable homogeneous state and multiple other states with various droplets.
This suggests that the inevitable surface tension in finite systems does not affect the multistability significantly.

\subsection*{Discussion}

We showed that multicomponent mixtures generally exhibit many metastable states with various phases, even if the homogeneous state is robust to fluctations.
While already binary mixture show a similar discrepancy in the form of the nucleation-and-growth regime, adding more components amplifies this behavior.
The fact that equilibrium states can exhibit many phases even when the homogeneous state is metastable further suggests that there are additional metastable states with fewer phases (\Figref{M-fig:metastable_energies}).
Since we only considered passive systems governed by equilibrium thermodynamics, these states must be separated by energy barriers, which could be overcome by nucleation.
We thus expect that starting from the homogeneous state several different types of droplets can be nucleated independently of each other, giving rise to a sequence of states with increasing droplet counts (\Figref{M-fig:multistability}).
Experimentalists begin to analyze such behaviors~\cite{erkamp2024Biomolecular}, but there are many open questions, including the number of metastable states, the magnitude of energy barriers, and whether there is a hierarchy of nucleation.

Our theory predicts two fundamental aspects of phase separation inside biological cells.
First, the fact that cells comprise millions of different components with weak interactions on the scale of the thermal energy~$k_\mathrm{B}T$ suggests that the equilibrium phase count is small compared to the component count (lower left corner of Fig.~\ref{M-fig:all_std_and_mean}B), which is consistent with the observed $10^1\ldots10^3$ different condensates~\cite{rostam2023CDCODE} that form from $10^5 \ldots 10^7$ different biomolecules~\cite{cooper2022Cell}.
Second, our theory implies multistability, which cells could exploit to create and remove different droplets independently of each other to fulfill separate functions robustly.
However, there is also the exciting possibility that formation of different droplets is coupled, which might be functionally relevant.
In both cases, cells could fine-tune the interactions of molecules, e.g., through evolutionary adaptation of the respective genes~\cite{zwicker2022Evolved}, even though our theory suggests that such fine-tuning is not necessary to obtain many multistable droplets.
Moreover, active processes would allow cells to exert finer control on when droplets form and how large they get~\cite{zwicker2025Physics,zwicker2022Intertwined,kirschbaum2021Controlling,soding2020Mechanisms,weber2019Physics}.
Taken together, our theory paints a picture of a multistable cytosol where hundred types of condensates can coexist independently.

We validated our scaling laws against numerical data, although the computational complexity limits the maximally attainable component count~$\Ncomp$.
While most scaling laws were derived based on analytical estimates, the dependence of the phase count $\barNphase$ on the interaction variance $\sigma_\chi^2$ was only obtained by analogy and might thus break down beyond the tested regime.
We here also focused on interactions that are randomly distributed with little structure, whereas realistic systems will likely have higher-order~\cite{luo2024Pairwise} and structured interactions~\cite{chen2023Programmable,qian2023Linking,kilgore2025Protein}, e.g., arising from evolutionary optimization, which affects phase separation~\cite{graf2022Thermodynamic,teixeira2023Liquid}.
Moreover, the composition of various components is typically broadly distributed, which might affect accessible phases.
To address these aspects, a detailed understanding of the geometry of phase space would be helpful, which certainly warrants further investigations and experimental validation.
Our results on the generic behavior of multicomponent mixtures will guide future research in these directions.

\bibliography{mm_binodal_refs_auto}

\onecolumngrid \appendix \setcounter{figure}{0} \renewcommand\thefigure{A\arabic{figure}} \renewcommand{\theHfigure}{A\arabic{figure}} \setcounter{equation}{0} \renewcommand\theequation{A\arabic{equation}} \renewcommand{\theHequation}{A\arabic{figure}}

\section{Simulation details}

\subsection{Stability of the homogeneous state}
\label{S-sec:hessian}
The stability of the homogeneous state is governed by the eigenvalues of the Hessian matrix $H_{ij}$ of the free energy given by \Eqrefst{M-eqn:fe_single_general} in the main text,
\begin{align}
    H_{ij} = \frac{\partial^2 f}{\partial \barphi_i \partial \barphi_j} = \chi_{ij} + \frac{\delta_{ij}}{\barphi_i} - \chi_{1j} - \chi_{i1} + \frac{1}{\barphi_1},
    \label{S-eqn:hessian_general}
\end{align}
with $i,j \in [2, \Ncomp]$.
Note that $\boldsymbol{H} = \{H_{ij}\}_{i,j=2}^\Ncomp$ is a $(\Ncomp-1)$-dimensional matrix since we removed the dependency on $\barphi_1$, using the incompressibility condition $\barphi_1 = 1 - \sum_{i=2}^\Ncomp \barphi_i$.
To obtain numerical estimates, we simply sample matrices and diagonalize them numerically.

\subsection{Numerical algorithm for finding coexisting phases in equilibrium}
\label{S-sec:numerics}
There are multiple challenges for determining coexisting phases in equilibrium in multicomponent mixture:
First, the optimization problem is high-dimensional.
Second, different types of constrains, such as incompressibility and volume conservation, need to be satisfied.
In addition, it is generally difficult to conclude whether the obtained phase-separated states with multiple coexisting phases are the true equilibrium or metastable states due to multistability.

In a previous publication~\cite{zwicker2022Evolved}, we designed an efficient algorithm to obtain coexisting phases by exchanging components between compartments, guided by thermodynamic properties such as osmotic pressure and chemical potential.
To achieve high performance, the constraint of volume conservation was relaxed, making it difficult to uniformly sample the entire phase diagram~\cite{zwicker2022Evolved}.
This flaw in sampling, as well as a small number of compartments, might have biased the result toward metastable states instead of the true equilibrium, which was highlighted in Ref.~\cite{jacobs2023Theory}.
To circumvent these challenges, we designed an improved method based on a free energy optimization strategy inspired by polymeric field theories~\cite{schmid1998Selfconsistentfield}, where the volume conservation is automatically guaranteed by introducing conjugate fields. 
Our method is now publicly available~\cite{qiang2025Flory}.

Instead of solving the balance equations of chemical potentials and osmotic pressures,  we adopt the more fundamental idea of free energy minimization, which is the origin of the balance equations.
The equilibrium coexisting phases can be obtained by optimizing the average free energy density given by \Eqrefst{M-eqn:fe_many} in the main text over all possible phase counts and phase compositions.
To allow different phase counts, we consider an ensemble of $M$ abstract compartments as proposed in Ref.~\cite{zwicker2022Evolved}, where $M$ is much larger than the number of components $\Ncomp$.
To alleviate the problem of negative volume fractions during the relaxation dynamics and conserve the average volume fractions, we extend the free energy of \Eqrefst{M-eqn:fe_many} into the form
\begin{align}
    \bar{f}
     & = \sum_{m=1}^M J_m \left[\frac{1}{2} \sum_{i,j=1}^\Ncomp \chi_{ij} \phi_i^{(m)} \phi_j^{(m)} - \sum_{i=1}^\Ncomp w_i^{(m)}\phi_i^{(m)} + \xi_m \biggl(\sum_{i=1}^\Ncomp \phi_i^{(m)}-1\biggr) \right] - \sum_{i=1}^{N}\bar{\phi_i}\ln Q_i +\eta\biggl(\sum_{m=1}^M J_m -1\biggr) ,
    \label{S-eqn:fe_unequal_grid}
\end{align}
with the single molecule partition function
\begin{align}
    Q_i = \sum_{m=1}^M J_m \exp\left(-w_i^{(m)}\right).
    \label{S-eqn:Q}
\end{align}
Here, $J_m$ are the relative volumes of the compartments, $w_i^{(m)}$ are the conjugate variables of $\phi_i^{(m)}$, and $\xi_m$ and $\eta$ are the Lagrangian multipliers for incompressibility of each compartment and compartment volume conservation, respectively.
Consequently, the extremum of \Eqref{S-eqn:fe_unequal_grid} with respect to $\xi(x)$ corresponds to incompressibility,
\begin{align}
    \frac{\partial \bar{f}}{\partial \xi_m} \propto \biggl(\sum_{i=1}^\Ncomp \phi_i^{(m)} - 1\biggr)J_m = 0 \quad \quad \Rightarrow  \quad \quad \sum_{i=1}^\Ncomp \phi_i^{(m)} = 1 \;,
    \label{S-eqn:incompressibility_unequal_grid}
\end{align}
the extremum with respect to $\eta$ corresponds to conservation of the total volume of all compartments,
\begin{align}
    \frac{\partial \bar{f}}{\partial \eta} \propto \sum_{m=1}^M J_m -1 = 0 \quad \quad \Rightarrow  \quad \quad \sum_{m=1}^M J_m = 1 \;,
    \label{S-eqn:measure_unequal_grid}
\end{align}
and the extremum with respect to $w_i^{(m)}$ defines the relationship between $\phi_i^{(m)}$ and $w_i^{(m)}$,
\begin{align}
    \frac{\partial \bar{f}}{\partial w_i^{(m)}} \propto -\phi_i^{(m)} J_m + \frac{\bar{\phi_i}}{Q_i}\exp\left(-w_i^{(m)} \right) J_m = 0  \quad \quad \Rightarrow  \quad \quad  \phi_i^{(m)} = \frac{\bar{\phi_i}}{Q_i}\exp\left(-w_i^{(m)}\right) .
    \label{S-eqn:volume_fraction_unequal_grid}
\end{align}
By inserting \Eqsref{S-eqn:incompressibility_unequal_grid}--\eqref{S-eqn:volume_fraction_unequal_grid} into \Eqref{S-eqn:fe_unequal_grid}, \Eqrefst{M-eqn:fe_many} is recovered except for a constant, which has no influences on thermodynamics.
Therefore, minimizing \Eqref{S-eqn:fe_unequal_grid} will naturally lead to balanced chemical potentials and osmotic pressures, and it is unnecessary to consider them explicitly.
To optimize the free energy density given by \Eqref{S-eqn:fe_unequal_grid}, we obtain the self-consistent equations
\begin{subequations}
    \label{S-eqn:sc_equations}
    \begin{align}
        1            & = \sum_{i=1}^\Ncomp \phi_i^{(m)}                                       \\
        1            & = \sum_{m=1}^M J_m                                                   \\
        \phi_i^{(m)} & = \frac{\bar{\phi_i}}{Q_i}\exp\left(-w_i^{(m)}\right)              \\
        w_i^{(m)}    & = \sum_{j=1}^\Ncomp \chi_{ij} \phi_j^{(m)} + \xi_m                     \\
        -\eta        & = \frac{1}{2}\sum_{i,j=1}^\Ncomp \chi_{ij} \phi_i^{(m)} \phi_j^{(m)}
        - \sum_{i=1}^\Ncomp w_i^{(m)}\phi_i^{(m)}
        + \xi_m \biggl(\sum_{i=1}^\Ncomp \phi_i^{(m)}-1\biggr)
        - \sum_{i=1} \phi_i^{(m)}\; .
    \end{align}
\end{subequations}
To solve these equations, we design the following iterative scheme
\begin{subequations}
    \label{S-eqn:sc_iteration}
    \begin{align}
        Q_i^{(m)}    & = \sum_{m=1}^M \exp\left(-w_i^{(m)}\right) J_m                                                            \\
        \phi_i^{(m)} & = \frac{\bar{\phi_i}}{Q_i^{(m)}}\exp\left(-w_i^{(m)}\right)                                             \\
        \xi_m        & = \frac{1}{\Ncomp} \left(\sum_{i=1}^{\Ncomp} w_i^{(m)} - \sum_{i,j=1}^{\Ncomp} \chi_{ij} \phi_j^{(m)} \right) \\
        \eta_m       & = -\frac{1}{2}\sum_{i,j=1}^\Ncomp \chi_{ij} \phi_i^{(m)} \phi_j^{(m)}
        + \sum_{i=1}^\Ncomp w_i^{(m)}\phi_i^{(m)} - \xi_m \biggl(\sum_{i=1}^\Ncomp \phi_i^{(m)}-1\biggr)
        + \sum_{i=1} \phi_i^{(m)}                                                                                                  \\
        \bar{\eta}   & = \sum_{m=1}^M  \eta_m J_m                                                                                \\
        w_i^{(m)*}   & = \sum_{j=1}^\Ncomp \chi_{ij} \phi_j^{(m)} + \xi^{(m)}                                                      \\
        J_m^*        & = J^{(m)} + \eta^{(m)} - \bar{\eta}\;,
    \end{align}
\end{subequations}
where the asterisks denote the output of the iteration.
In order to improve numerical stability, we also adopt the simple mixing strategy,
\begin{subequations}
    \begin{align}
        w_i^{(m),\mathrm{new}} & = w_i^{(m)} + \alpha \left(w_i^{(m)*} - w_i^{(m)}\right) \\
        J^{(m),\mathrm{new}}   & = J_m + \beta \left(J_m^* - J_m\right)\;,
    \end{align}
\end{subequations}
where $\alpha$ and $\beta$ are two empirical constants, which are usually chosen near $10^{-3}$.
We note again that in such iteration scheme the problem of negative volume fractions is relieved.
However, there is no guarantee that relative compartment volume $J_m$ is always positive.
Although the algorithm does not suffer from negative $J_m$, negative $J_m$ implies that the system might be outside of the allowed region on the tie hyperplane.
To alleviate this, we always use $\beta$ smaller than $\alpha$, and adopt a killing-and-revive strategy to correct the worst cases:
Once $J_m$ is found to be negative at certain $m$, e.g. $m_0$, the corresponding compartment is considered ``dead'' and is going to be revived by resetting $J_{m_0}$ to its initial value, and the corresponding $w_i^{(m_0)}$ will be redrawn from random distributions.
To obey volume conservation, all other $J_m$ will be renormalized.
The same scheme is used to initialize the simulation, i.e., all compartments are considered ``dead'' at the beginning of the simulation.

Due to multistability, this algorithm does not guarantee that the true equilibrium state is always found.
Since pairwise exchange of volume fractions is replaced by a global redistribution, the new numerical method has lower time complexity with respect to the number of compartments $M$ compared to the method in Ref. \cite{zwicker2022Evolved}.
Therefore we handle the problem of multistability by launching many more compartments than the number of components, $M\gg\Ncomp$.
In all our numerical results, the number of compartments are at least $M=16\Ncomp$.
We justified this choice by increasing number of compartments until both the number and the compositions of unique coexisting phases do not change.
Under such setting, the equilibrium coexisting phases are prominently obtained.

\subsection{Continuous field theory}
The continuous simulations shown in \Figref{M-fig:multistability} of the main text are based on a spatially averaged form of the non-dimensional free energy density in \Eqref{M-eqn:fe_many}, extended with the interfacial energy to penalize sharp interfaces,
\begin{align}
    \bar{f}_\mathrm{spatial} = \frac{1}{V} \int \left[\frac12\sum_{i,j=1}^\Ncomp\chi_{ij} \phi_i \phi_j + \sum_{i=1}^\Ncomp \phi_i \ln \phi_i + \frac{1}{2}\ell^2 \sum_{i=1}^\Ncomp \left|\boldsymbol{\nabla}\phi_i\right|^2 \right] \mathrm{d} \boldsymbol{r} \;,
\end{align}
where $V$ is the volume of the system and $\ell$ is the interfacial width.
The evolution of the conserved volume fraction fields $\phi_i$ then follows the Cahn-Hilliard equation
\begin{align}
    \frac{\partial \phi_i}{\partial t} = \boldsymbol{\nabla} \cdot \sum_{j=1}^\Ncomp \Lambda_{ij} \boldsymbol{\nabla} \mu_j \;,
    \label{S-eqn:dynamical_model}
\end{align}
where $\mu_i = V \delta \bar{f}_\mathrm{spatial}/\delta \phi_i$ is the non-dimensional chemical potential of each component, and 
$\Lambda_{ij}$ is the mobility matrix selected to satisfy the incompressibility $\sum_i \phi_i = 1$~\cite{zwicker2025Physics}.
Specifically, we choose
\begin{align}
    \Lambda_{ij} = \lambda\left(\phi_i \delta_{ij} - \phi_i \phi_j\right) \;.
\end{align}
The mobility matrix has no influence on coexisting phases in the stationary state, while the interfacial term proportional to $\ell$ only mildly modifies the coexisting concentrations via Laplace pressure effects~\cite{zwicker2025Physics}.

We obtained the metastable configurations in \Figref{M-fig:multistability} by numerically solving \Eqref{S-eqn:dynamical_model} from various initial configurations.
We generate initial configurations by adding random noise to the homogeneous state followed by adjustments to satisfy the incompressibility. 
A semi-implicit method is used to improve the numerical stability.
For all the simulations, we use $\ell=4$, $\lambda=1$ and a periodic 2D simulation box of $64 \times 64$.

\section{Identical interactions and symmetric composition}
In this section, we first derive the stability condition for mixtures with identical interactions, $\chi_{ij}=\chimean(1-\delta_{ij})$, and symmetric composition, $\barphi_i=1/\Ncomp$.
Then in \ref{S-sec:eq_cond}, we provide a sufficient condition for the equilibrium state with multiple coexisting phases.
In \ref{S-sec:dilute_phi}, we derive the analytical approximation of the volume fraction of the dilute component in each of the coexisting phases. 
In \ref{S-sec:enrich_justification}, we justify the assumption for the compositions of the coexisting phases in \ref{S-sec:eq_cond} for completeness.

\subsection{Stability analysis of homogeneous states}
\label{S-sec:unstable_cond}
\begin{figure}[t]
    \begin{center}
        \includegraphics[width = 0.5 \linewidth]{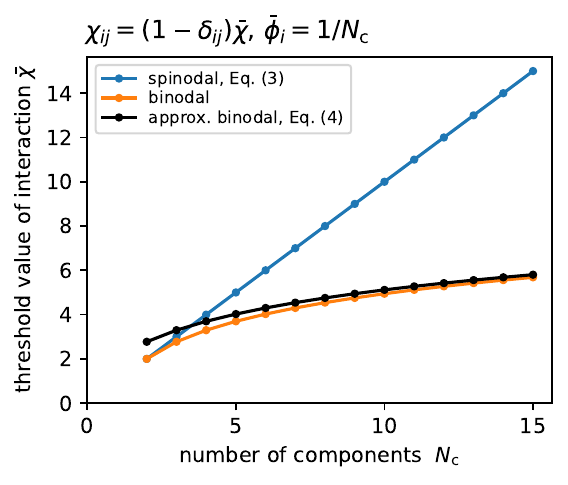}
    \end{center}
    \caption{
        \textbf{Interaction strength for stable multiphase coexistence differs from stability analysis in mixtures with identical interactions and symmetric composition.}
        A homogeneous mixture is unstable for interactions~$\chimean$ above the spinodal (blue line, given by \Eqrefst{M-eqn:spinodal_symchi_symphi} in the main text).
        In contrast, phase separation is energetically favorable for $\chimean$ above the binodal (orange line, obtained numerically).
        The black line shows the analytical approximation given by \Eqrefst{M-eqn:binodal_symchi_symphi_approx} in the main text.
    }
    \label{S-fig:symmetric}
\end{figure}

With identical interactions and symmetric composition, the Hessian given by \Eqref{S-eqn:hessian_general} reads 
   $ H_{ij} =(\delta_{ij}+1) (\Ncomp - \chi )$,
which has two unique eigenvalues $\Ncomp - \chi$ and $\Ncomp(\Ncomp - \chi)$.
The homogeneous state is unstable when the Hessian above is no longer positive definite, i.e., when $\chimean>\Ncomp$; 
see the blue line in \Figref{S-fig:symmetric}.

\subsection{Sufficient condition for the equilibrium phase-separated state}
\label{S-sec:eq_cond}
To obtain the sufficient condition for the state with multiple coexisting phases to be stable, we first derive the free energy of the homogeneous state.
With identical interactions and symmetric composition, the free energy density given by \Eqrefst{M-eqn:fe_single_general} reads
\begin{align}
    f = \frac{\Ncomp-1}{2\Ncomp}\chimean -\ln \Ncomp
    \;.
    \label{S-eqn:fe_single_symchi_symphi}
\end{align}
Since all components interact equally, coexisting phases must have permutational symmetry.
Each phase then enriches one component with volume fraction $\phi_\mathrm{h} = 1-(\Ncomp-1) \phi_\mathrm{l}$, while all other components have the same volume fraction $\phi_\mathrm{l}$.
In principle, there could also be coexisting phases that enrich more components per phase, but we will show in \ref{S-sec:enrich_justification} that the state above is the energetically preferred phase-separated state when $\bar\chi$ is increased.
For this state, the free energy density in each phase reads
\begin{align}
    f(\phi_\mathrm{l})
     & =
    (\Ncomp-1)\chimean\phi_\mathrm{h}\phi_\mathrm{l} + \frac{(\Ncomp-1)(\Ncomp-2)}{2} \chimean \phi_\mathrm{l}^2
    + \phi_\mathrm{h} \ln \phi_\mathrm{h} + (\Ncomp-1)\phi_\mathrm{l} \ln \phi_\mathrm{l}
    \;.
    \label{S-eqn:fe_single_symchi_oneenriched}
\end{align}
Note that \Eqref{S-eqn:fe_single_symchi_oneenriched} reduces to \Eqref{S-eqn:fe_single_symchi_symphi} in the homogeneous case when $\phi_\mathrm{h}=\phi_\mathrm{l}=1/\Ncomp$.
Since all coexisting phases have the same free energy density due to permutational symmetry, the coexisting phases are energetically favored over the homogeneous state if the minimal value of \Eqref{S-eqn:fe_single_symchi_oneenriched} is lower than that of the homogeneous state given by \Eqref{S-eqn:fe_single_symchi_symphi}.
The orange curve in \Figref{S-fig:symmetric} shows the results from a numerical minimization of \Eqref{S-eqn:fe_single_symchi_oneenriched} compared to \Eqref{S-eqn:fe_single_symchi_symphi}.
To obtain an analytical approximation of the minimal interaction strength necessary for phase separation, we note that with $\phi_\mathrm{l}\rightarrow +0$, the free energy $f(\phi_\mathrm{l}) \rightarrow 0$, while its derivative $\dd f(\phi_\mathrm{l}) / \dd \phi_\mathrm{l} \rightarrow -\infty$.
Using the mean value theorem, we then conclude that \Eqref{S-eqn:fe_single_symchi_oneenriched} will have a minimum lower than \Eqref{S-eqn:fe_single_symchi_symphi} when $f(1/\Ncomp)$ is positive, implying
\begin{align}
    \chimean > \frac{2\Ncomp}{\Ncomp-1}\ln \Ncomp
    \;,
    \label{S-eqn:binodal_symchi_symphi_approx}
\end{align}
which is \Eqrefst{M-eqn:binodal_symchi_symphi_approx} in the main text.

\subsection{Volume fraction of the dilute component in coexisting phases}
\label{S-sec:dilute_phi}
We further consider the volume fraction $\phi_\mathrm{l}$ of the dilute components in coexisting phases.
Denoting by  $\phi_\mathrm{l} = \phi^-$ the minimizer of \Eqref{S-eqn:fe_single_symchi_oneenriched}, $\phi^-$ satisfies the equation
\begin{align}
    \chimean = \frac{1}{1-\Ncomp \phi^-}\biggl( \ln \left[ 1- (\Ncomp-1)\phi^- \right] -  \ln \phi^- \biggr)
    \;.
    \label{S-eqn:phi_star_equation}
\end{align}
Assuming that $\phi^-$ is much smaller that $1/\Ncomp$, we obtain
\begin{align}
    \phi^-(\Ncomp) \approx - \frac{W_0\left[-e^{-\chimean}\left(\Ncomp(\chimean -1) + 1\right)\right]}{\Ncomp(\chimean -1) + 1}
    \;,
    \label{S-eqn:phi_star_approx_1}
\end{align}
where $W_0(r)$ is the principal branch of the Lambert $W$ function, which satisfies $W(r)e^{W(r)}=r$.
When $\chimean \gg \ln \Ncomp$, the equation above can be further simplified to
\begin{align}
    \phi^- \overset{\chimean \gg 1 }{\approx} e^{-\chimean}
    \;,
    \label{S-eqn:phi_star_approx_2}
\end{align}
which is independent of $\Ncomp$.
Note that \Eqref{S-eqn:phi_star_approx_2} also shows that the assumption $\phi^- \ll 1/\Ncomp$ is valid when $\chimean \gg \ln \Ncomp$.

\subsection{Justification for each phase enriching one component}
\label{S-sec:enrich_justification}
Here we show that with $\chimean \gg \ln \Ncomp$, a system with identical interactions and symmetric composition will minimize its free energy by separating into phases where each phase concentrates one component.
The argument comprises three steps:

\paragraph{The coexisting phases in equilibrium must have the same free energy density.} 
This can be proved by contradiction:
Assuming coexisting phases have different free energy energies in equilibrium, there must exist one of the coexisting phases with the lowest free energy density.
Suppose such phase has a composition denoted by the vector $\boldsymbol{\phi}=[\phi_1, \phi_2 \hdots \phi_\Ncomp]$, then a new collection of phases can be constructed by permuting the entries of the vector.
Since all the interactions are identical, these phases share the same free energy density.
Also the average composition of these phases is symmetric, meaning that the coexistence of all the phases in the collection becomes a possible candidate state for the equilibrium state of the mixture.
Since this state has lower free energy density than the assumed state, the initial assumption of equilibrium is violated.
Therefore, with identical interactions and symmetric composition, all coexisting phases in equilibrium must have the same free energy density.

\paragraph{Each component can either be dilute/concentrated in each of the coexisting phases.}
As a result of equal free energy density, the composition of the coexisting phases must be the local minimizer of the free energy function, otherwise a new set of coexisting phases with lower free energy can be found.
Therefore, the compositions must satisfy ${\partial f}/{\partial \phi_i}=0$ for $i=2 \hdots \Ncomp$, which is the case for
\begin{align}
    -\chimean \phi_i + \ln \phi_i = \mathrm{constant}
    \label{S-eqn:phi_star_condition}
\end{align}
for $i=1,2\hdots \Ncomp$, where the constant is the same across different components $i$.
Since the function $g(\phi) = -\chimean \phi + \ln \phi$ can at most have two piecewise monotonic domains in $\phi\in(0,1)$, the equation above can have at most two solutions, meaning that each component can either be dilute or concentrated, while all the dilute components share the same composition $\phi_\mathrm{l}$ and all the concentrated components share the same composition $\phi_\mathrm{h}$.

\paragraph{Each of the coexisting phases prefers to concentrate only one component.}
Suppose the system would phase separate into phases that concentrate $K$ components with volume fraction $\phi_\mathrm{h}$, then in each phase $\Ncomp-K$ components are dilute and share the volume fraction $\phi_\mathrm{l}$, satisfying $K \phi_\mathrm{h} + (\Ncomp-K)\phi_\mathrm{l} = 1$.
Using \Eqref{S-eqn:phi_star_condition}, we denote by $\phi_\mathrm{l} = \phi^-(\Ncomp)$ the volume fraction that minimizes the free energy, which satisfies
\begin{align}
    \chimean = \frac{\ln \phi_\mathrm{h} -  \ln \phi_\mathrm{l}}{\phi_\mathrm{h} - \phi_\mathrm{l}}
    \;.
    \label{S-eqn:phi_star_equation_ex}
\end{align}
In the limit $\phi_\mathrm{l}\ll 1$, we find
\begin{align}
    \phi_\mathrm{l} \approx \frac{1}{K}e^{-K \chimean}
    \;.
    \label{S-eqn:phi_star_approx_2_ex}
\end{align}
Since the right hand side decreases monotonically with $K$ for $\chimean>0$, the dilute components become more dilute for larger~$K$.
For sufficient large value of $\chimean$, the concentrated components will take over the vast majority of the phase ($\phi_\mathrm{h} \approx K^{-1}$), making these concentrated components effectively a subsystem of $K$ components with equal interactions and volume fractions.
This subsystem will undergo phase separation again since the homogeneous state is more prone to phase separation for fewer components; see \Figref{S-fig:symmetric}.
Taken together, the system prefers to enrich one component in each phase since the enriched components would otherwise phase separate from each other.

Combining the three steps above, in the case of identical interactions and symmetric average composition, we find that each of the coexisting phases must enrich exactly one component when the phase-separated state is the equilibrium state.
Intuitively, all components dislike each other equally and they prefer to separate from each other when possible.
Phases that enrich more components are not energetically preferred without other effects such as higher-ordered interactions~\cite{luo2024Pairwise}.

\section{Identical interactions and uniformly distributed composition}

\subsection{Average number of unstable modes}
\label{S-sec:nu_average}
We here consider the number of unstable modes averaged over the entire phase diagram (random average composition satisfying $\sum_i \barphi_i=1$) for identical interactions, $\chi_{ij} = \chimean(1-\delta_{ij})$.
The Hessian given by \Eqref{S-eqn:hessian_general} can be simplified and rewritten in matrix form,
\begin{align}
    \boldsymbol{H} = \mathrm{diag}\left\{\frac{1}{\barphi_i} - \chi \right\}_{i=2}^\Ncomp + \left(\frac{1}{\barphi_1} - \chi\right) \boldsymbol{u}\boldsymbol{u}^T
    \;,
    \label{S-eqn:hessian_symchi}
\end{align}
where $\boldsymbol{u}$ is an all-one vector in $\Ncomp - 1$ dimensions.
The rank-$n$ leading minor of $\boldsymbol{H}$, denoted by $\boldsymbol{H}_n$, reads
\begin{align}
    \boldsymbol{H}_n = \mathrm{diag} \left\{\frac{1}{\barphi_i} - \chi  \right\}_{i=2}^{n+1} +  \left(\frac{1}{\barphi_1} - \chi \right) \boldsymbol{u}_n\boldsymbol{u}_n^T
    \;,
    \label{S-eqn:hessian_minor}
\end{align}
with $n=1,2\hdots\Ncomp -1$ and $\boldsymbol{u}_n$ is a all-one vector in $n$ dimensions.
Making use of the Cauchy interlacing theorem, the number of negative eigenvalues of $\boldsymbol{H}$ can be obtained by calculating the number of sign changes in the sequence
\begin{align}
    \det{\boldsymbol{H}_0},\det{\boldsymbol{H}_1}, \hdots, \det{\boldsymbol{H}_{\Ncomp -1}}
    \;,
    \label{S-eqn:hessian_minor_sequence}
\end{align}
where $\det{\boldsymbol{H}_0}$ is defined to be 1.
The terms $\det{\boldsymbol{H}_n}$ can be determined using the Weinstein-Aronszajn identity,
\begin{align}
    \det{\boldsymbol{H}_n}
     & =  \left(\frac{1}{\barphi_1} - \chi \right)^{n} \det \mathrm{diag}  \left\{ \frac{\barphi_i^{-1} - \chi}{\barphi_1^{-1} - \chi}  \right\}_{i=2}^{n+1} \det  \left( \boldsymbol{I} + \boldsymbol{u}_n\boldsymbol{u}_n^T \mathrm{diag} \left\{ \frac{\barphi_1^{-1} - \chi}{\barphi_i^{-1} - \chi}  \right\}_{i=2}^{n+1} \right) \notag \\
     & =\prod_{i=1}^{n+1}  \left(\frac{1}{\barphi_i} - \chi \right) \left(\sum _{i=1}^{n+1} \frac{1}{\frac{1}{\barphi_i} - \chi} \right).
    \label{S-eqn:hessian_minor_det}
\end{align}
We here assume that $\barphi_i^{-1} \neq \chi$ since we are interested in the average behavior over the entire phase diagram, where the points violating this condition have zero measure.
Since the number of negative eigenvalues is invariant if the matrix is permuted by both rows and columns at the same time, we can also require that sequence
\begin{align}
    \frac{1}{\barphi_1} - \chi,\frac{1}{\barphi_2} - \chi, \hdots, \frac{1}{\barphi_\Ncomp} - \chi
    \label{S-eqn:phi_sequence}
\end{align}
decreases monotonically.
Comparing \Eqref{S-eqn:hessian_minor_sequence} and \Eqref{S-eqn:hessian_minor_det} with \Seqref{S-eqn:phi_sequence}, we obtain that the number of sign changes in \Seqref{S-eqn:hessian_minor_sequence} can only be either (i) the number of negative elements in \Seqref{S-eqn:phi_sequence}, denoted by $N_-$, or (ii) one less, $N_--1$, since the term $\sum _{i=1}^\Ncomp 1/(\barphi_i^{-1} - \chi)$ can at most change the sign once itself and cancel the sign changes of $\prod_{i=1}^\Ncomp (\barphi_i^{-1} - \chi)$ once.
We thus obtain the relationship between $\Nunstable$ and $N_-$ given in the main text.

\subsection{Bounds for number of coexisting phases with identical interactions}
\label{S-sec:np_bounds}
In the main text, the average number of coexisting phases over the entire phase diagram, $\barNphase$, was obtained by constructing  lower and upper bounds.
For $\Ncomp =3$, the construction is visualized in \Figrefst{M-fig:binodal_sym} of the main text, and we now explain this construction for general $\Ncomp$.
For identical interactions, a state with $K$ coexisting phases necessarily has $\Ncomp-K$ non-phase-separating components (for the same reasons as discussed in section \ref{S-sec:enrich_justification}).
The composition of these non-phase-separating components will be the same in all $K$ coexisting phases, again due to identical interactions.
The contribution of these components to the free energy density is thus identical in all phases, implying that these non-phase-separating components are effectively inert.
The geometric interpretation is that all $K$-dimensional tie hyperplanes are parallel to one of the $K$-dimensional boundaries of the entire phase diagram.
Assuming the total volume fractions of all non-phase-separating component are $\Delta \phi$, similar to \Eqref{S-eqn:phi_star_equation_ex}, we still obtain
\begin{align}
    \chimean = \frac{\ln \phi_\mathrm{h} -  \ln \phi_\mathrm{l}}{\phi_\mathrm{h} - \phi_\mathrm{l}}
    \;,
    \label{S-eqn:phi_star_equation_ex_2}
\end{align}
where $\phi_\mathrm{h} = 1 - \Delta \phi - (K-1) \phi_\mathrm{l}$ is the volume fraction of the concentrated component of one phase, whereas $\phi_\mathrm{l}$ is the volume fraction of all $K-1$ components that are diluted in this phase.
Again assuming $\phi_\mathrm{l} \ll 1/\Ncomp$, we obtain the approximate solution
\begin{align}
    \phi_\mathrm{l} \approx - \frac{W_0\left[-e^{\Delta \phi (\chimean-1)}e^{-\chimean}\left(K(\chimean -1) + 1\right)\right]}{K(\chimean -1) + 1}
    \;.
    \label{S-eqn:phi_star_ex_approx}
\end{align}
Since $W_0$ is a monotonic increasing function, $\phi_\mathrm{l}$ increases with larger $\Delta \phi$.
This implies that increasing the volume fraction of non-phase-separating components disfavors phase separation, leading to higher concentration~$\phi_\mathrm{l}$ of the dilute component and lower concentration~$\phi_\mathrm{h}$ of the concentrated component in coexisting phases.
Based on these monotonic dependencies, we construct upper bounds and lower bounds using the minimal and  maximal value of $\phi_\mathrm{l}$ of the real phase boundaries, which correspond to the outermost point and the innermost point of the phase boundaries in the real phase diagram.

\section{Random interactions}

\subsection{Governing parameters for average phase count with random interactions}
\label{S-sec:np_control_para}
In the case of identical interactions, we obtained analytical governing parameters and asymptotic rules for the average number of unstable modes, $\barNunstable$, and average number of coexisting phases, $\barNphase$; see main text.
In contrast, for random interactions with zero mean, random matrix theories only grants us access to the governing parameter for $\barNunstable$, but it does not directly reveal anything about $\barNphase$.
To infer the governing parameter for $\barNphase$, we exploit connections between the results from the other three known cases.
For simplicity, we assume vanishing mean value of the random interactions  in this section, unless specified otherwise.

To predict the scaling of $\barNphase$ with $\Ncomp$ and $\chistd$, we first recall the results for $\barNunstable$ for random interactions.
Ignoring the sole outlier in the spectrum of the Hessian matrix given by \Eqref{S-eqn:hessian_general}, the average number of negative eigenvalues is approximately the number of eigenvalues of the interaction matrix $\chi_{ij}$ smaller than $-1/\barphi_i=-\Ncomp$, when $\barphi_i = 1/\Ncomp$.
In the case of asymmetric composition, the eigenvalue spectrum will be expanded, but in the limit of larger interaction variance, the shape of the spectrum is still dominated by the random interaction matrix~\cite{thewes2023Composition}.
In addition, we expect that this spectrum expansion has no preference for positive or negative values and thus averages out when we consider the average behavior over the entire phase diagram $\sum_i \barphi_i=1$.
Therefore, to find the governing parameters, we check the situation of symmetric composition and hypothesize that the result generalizes to the average behavior over the entire phase diagram.
Denoting the eigenvalues of $\chi_{ij}$ by $\lambda_k$ with $k=1,2\hdots\Ncomp-1$, the number of unstable modes, $\barNunstable$, can be roughly estimated by
\begin{align}
    \barNunstable \approx E_{\Ncomp}[\lambda_k > \barphi_i^{-1}] \quad \text{for} \quad \chi_{ij} \sim \mathcal{N}(0, \chistd)\;.
    \label{S-eqn:unstable_random_detailed}
\end{align}
Since $\lambda_k$ is distributed according to a semicircle law with radius $2\chistd \sqrt{\Ncomp}$, we set $\lambda_k = 2\chistd \sqrt{\Ncomp} \gamma$, where $\gamma$ is a random variable according to a semicircle law with unit radius.
We thus obtain $\barNunstable \approx E_{\Ncomp}[2\chistd \sqrt{\Ncomp} \gamma > \barphi_i^{-1}] = E_{\Ncomp}[\gamma> \sqrt{\Ncomp}/\chistd]$, indicating that $\barNunstable$ is governed by the $\chistd/\sqrt{\Ncomp}$ when $\chi_{ij} \sim \mathcal{N}(0, \chistd)$.
However, even though this derivation is only approximative, the literature~\cite{thewes2023Composition} and our numerical data shown in \Figrefst{M-fig:binodal_random}(a) of the main text suggest that the governing parameter is still valid when $\barNunstable$ is obtained for a uniform average over the entire phase diagram.

Although \Eqref{S-eqn:unstable_random_detailed} provides no direct information about $\barNphase$, it strikingly shares the same form as the relations for $\barNunstable$ for identical interactions $\chi_{ij} \sim \mathcal{N}(\chimean, 0)$, allowing us to build a connection with $\barNphase$ by analogy.
From \Eqrefst{M-eqn:unstable_modes_count_asymptotic} and \Eqrefst{M-eqn:phase_count_asymptotic} in the main text, we obtain
\begin{subequations}
    \begin{align}
        \barNunstable & \approx E_{\Ncomp}[\barphi_i>\chimean^{-1}] =E_{\Ncomp}[ \chimean >\barphi_i^{-1}] \quad \text{for} \quad \chi_{ij} \sim \mathcal{N}(\chimean, 0)              \\
        \barNphase    & \approx E_{\Ncomp}[\barphi_i>e^{-\chimean}] =E_{\Ncomp}[e^\chimean>\barphi_i^{-1}] \quad \text{for} \quad \chi_{ij} \sim \mathcal{N}(\chimean, 0)\;.
    \end{align}
\end{subequations}
Note that $\chimean$ is tightly related to the eigenvalues, since in the case of identical interactions the relevant eigenvalues are all equal to $-\chimean$.
The only difference between the unstable modes and the phase count in the equations above is that the interaction $\chimean$ is modified to $e^\chimean$.
Inspired by this, we speculate that such a modification can also be applied to the case of random interactions, $\chi_{ij} \approx \mathcal{N}(0, \chistd)$.
We thus propose
\begin{align}
    \barNphase \approx E_{\Ncomp}[e^{\lambda_k}>\barphi_i^{-1}] \quad \text{for} \quad \chi_{ij} \sim \mathcal{N}(0, \chistd) \;.
\end{align}
Since $\lambda_k = 2\chistd \sqrt{\Ncomp} \gamma$, and the average value of $\barphi_i$ is $1/\Ncomp$, we further propose that $\barNphase \approx E_{\Ncomp}[2\chistd \sqrt{\Ncomp} \gamma > \ln \Ncomp]$. 
It leads to the governing parameter $\chistd \sqrt{\Ncomp}/\ln \Ncomp$, which we verified numerically in the main text.

\subsection{Verification of the numerically estimated master functions}
\label{S-sec:master_function}
\begin{figure}[t]
    \begin{center}
        \includegraphics[width = 0.8 \linewidth]{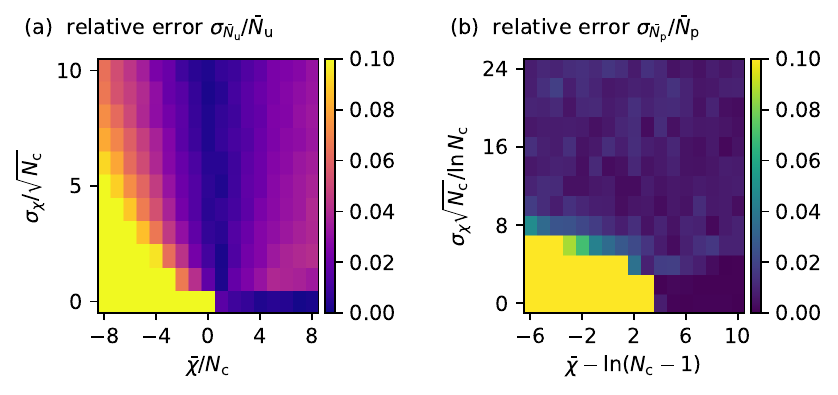}
    \end{center}
    \caption{
        \textbf{Relative standard deviation of the numerically estimated master functions for $\Ncomp \in [16,32]$.}
        (a) Relative standard deviation of number of unstable modes, $\sigma_{\barNunstable}/\barNunstable$, and (b) relative standard deviation of number of coexisting phases, $\sigma_{\barNphase}/\barNphase$, as a function of two scaling variables (a) $\chimean/\Ncomp$ and $\chistd/\sqrt{\Ncomp}$, and (b) $\chimean-\ln(\Ncomp-1)$ and $\chistd\sqrt{\Ncomp}/\ln{\Ncomp}$, respectively.
        Note that the relative error in the left corner of both panels is high while the absolute error is still low, since the reference values $\barNunstable \approx 0$ and $\barNphase \approx 1$ are very low in this region.
        Standard deviations are calculated for $\Ncomp=16,20,24,28,32$.
    }
    \label{S-fig:all_std}
\end{figure}

\begin{figure}[t]
    \begin{center}
        \includegraphics[width = 0.8 \linewidth]{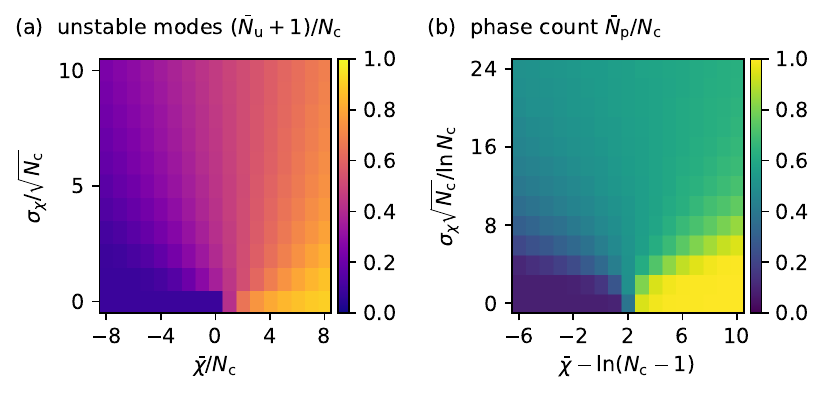}
    \end{center}
    \caption{
        \textbf{Numerically estimated master functions for $\Ncomp \in [8,16]$.}
        Data are averaged over $\Ncomp=8,10,12,14,16$.
        300-3000 random pairs of interaction matrices and compositions are drawn for each fixed $\Ncomp$, $\chimean$ and $\chistd$.
        Other elements are the same as \Figrefst{M-fig:all_std_and_mean} in the main text.
    }
    \label{S-fig:all_mean_lower}
\end{figure}
\begin{figure}[t]
    \begin{center}
        \includegraphics[width = 0.8 \linewidth]{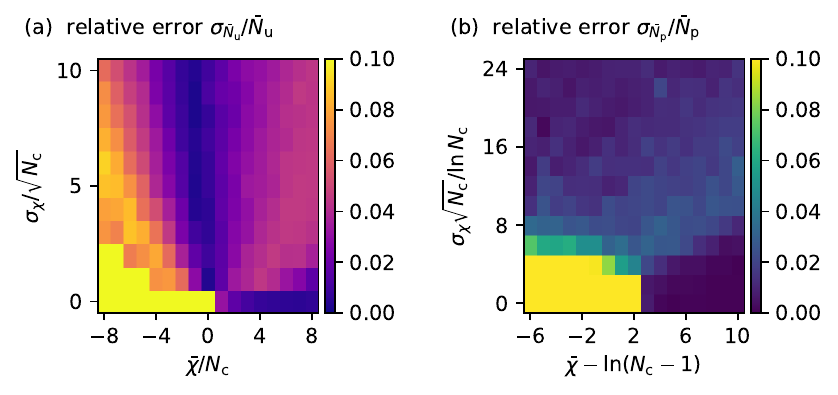}
    \end{center}
    \caption{
        \textbf{Relative standard deviation of the numerically estimated of master functions for $\Ncomp \in [8,16]$.}
        Standard deviations are calculated for $\Ncomp=8,10,12,14,16$.
        300-3000 samples of random interaction matrices and compositions are drawn for each fixed $\Ncomp$, $\chimean$ and $\chistd$.
        Other elements are the same as \Figref{S-fig:all_std}.
    }
    \label{S-fig:all_std_lower}
\end{figure}

To confirm the scalings proposed in \Eqrefst{M-eqn:master_2D} in the main text, and the associated master functions shown in \Figrefst{M-fig:all_std_and_mean}, we calculate the relative standard deviations across the selected number of components.
\Figref{S-fig:all_std}a shows that the deviation between the true values of  $(\barNunstable + 1)/\Ncomp$ and the values predicted by the master function $g_\mathrm{u}$ is below $5\%$ in the relevant parameter regime of repulsive interactions ($\chimean>0$).
\Figref{S-fig:all_std}b shows that the deviation is even smaller for the phase count $\barNphase/\Ncomp$, showing that it is incredibly well explained by the master function $g_\mathrm{p}$ for positive $\chimean$.
In both datasets, the relative deviation is higher in the lower left corner of the plot ($\chimean < 0$ and small $\chistd$).
This is expected since most cases exhibit a homogeneous system, so the number of unstable modes is close to $0$, while number of coexisting phases is close to $1$.
In these cases, the relative standard deviation degenerates to the standard deviation of the $1/\Ncomp$, which is apparently high with respect to $1/\Ncomp$ itself.

To validate our result further, we repeat the same procedure for determining the master functions using half the component count $\Ncomp$ everywhere.
The respective deviations are very similar (compare  \Figref{S-fig:all_std} and \Figref{S-fig:all_std_lower}), and comparing
\Figref{S-fig:all_mean_lower} of the main text to \Figrefst{M-fig:all_std_and_mean} shows that the resulting master functions are very similar.
Taken together, this suggests that the master functions shown in  \Figref{S-fig:all_mean_lower} are reliable.

\subsection{Linear regime of average unstable modes/coexisting phases count with respect to mean interaction}
\label{S-sec:linear_regime}
\begin{figure}[t]
    \begin{center}
        \includegraphics[width = 0.8 \linewidth]{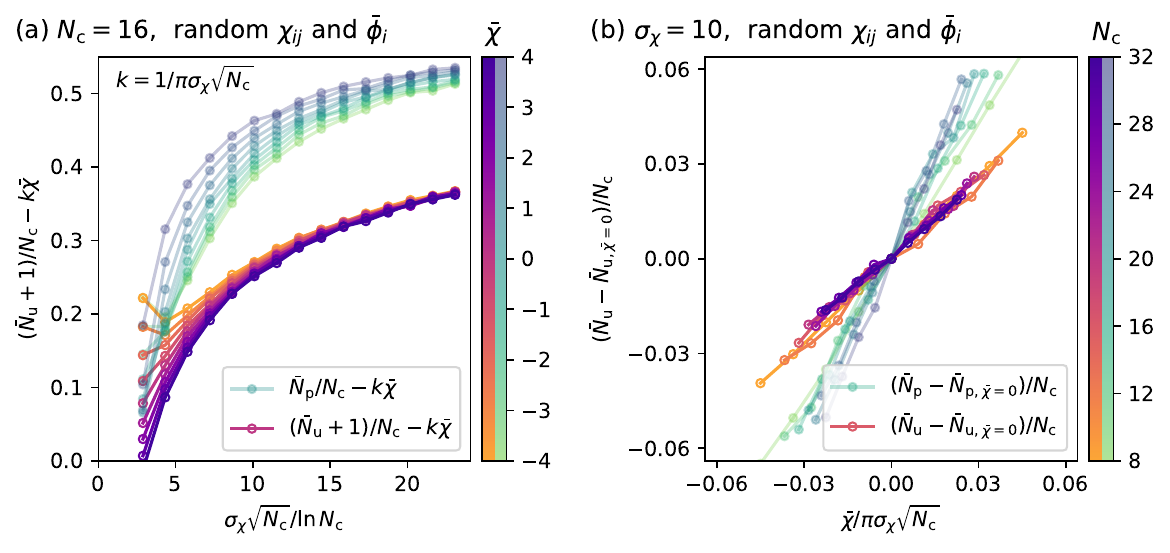}
    \end{center}
    \caption{
        \textbf{Linear correction proportional to $\chimean$ collapses data of the number of unstable modes.}
        (a) Shifted normalized average number of unstable modes (orange-purple) and coexisting phases (green-blue) as a function of the scaled standard deviation of the interactions for various mean interactions $\chimean$ at fixed component count $\Ncomp=16$. The shift coefficient $k=1/\pi \chistd \sqrt{\Ncomp}$.
        (b) Changes in the normalized average number of unstable modes (orange-purple) and coexisting phases (green-blue) as a function of the scaled mean interactions for various component counts $\Ncomp$ at fixed standard deviation of the interactions, $\chistd=10$.
    }
    \label{S-fig:linear_spinodal}
\end{figure}

\begin{figure}[t]
    \begin{center}
        \includegraphics[width = 0.8 \linewidth]{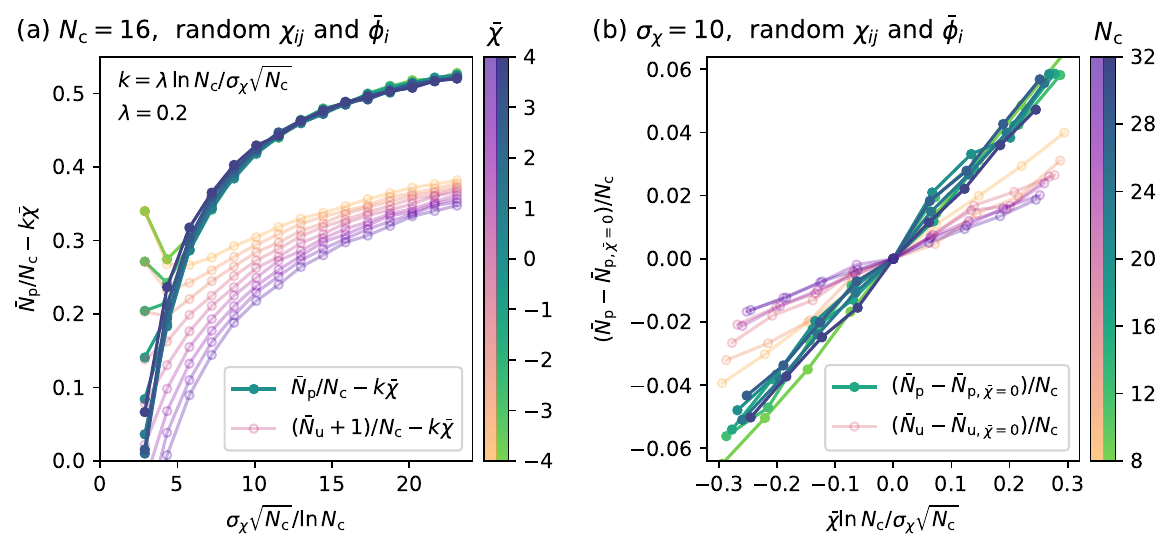}
    \end{center}
    \caption{
        \textbf{Linear correction proportional to $\chimean$ collapses data of the number of coexisting phases.}
        (a) Shifted normalized average number of unstable modes (orange-purple) and coexisting phases (green-blue) as a function of the scaled standard deviation of the interactions for various mean interactions $\chimean$ at fixed component count $\Ncomp=16$. The shift coefficient $k=\lambda \ln \Ncomp/\chistd \sqrt{\Ncomp}$, where $\lambda$ is a fitting constant independent of $\Ncomp$.
        (b) Changes in the normalized average number of unstable modes (orange-purple) and coexisting phases (green-blue) as a function of the scaled mean interactions for various component counts $\Ncomp$ at fixed standard deviation of the interactions, $\chistd=10$.
    }
    \label{S-fig:linear_binodal}
\end{figure}

The master functions $g_\mathrm{u}$ and $g_\mathrm{p}$ shown in  \Figref{S-fig:all_mean_lower} in the main text were only estimated numerically.
However, their smooth behavior in the region of large $\chistd$ and small $\chimean$ suggests that there is a simpler relationship in this region.
To obtain such a relationship, we investigate deviations from the line for $\chimean=0$ for small $|\chimean|$.

For the unstable modes, $(\barNunstable+1)/\Ncomp$, such relationship can be inferred again from random matrix theory.
The eigenvalues of the random interaction matrix are distributed according to the semicircle law with radius $r = 2 \chistd \sqrt{\Ncomp}$.
Adding a mean value $\chimean$ shifts the distribution by $-\chimean$ accordingly.
Therefore, the normalized number of unstable modes will increase roughly by $2 r \chimean / (\pi r^2) = \chimean/(\pi \chistd \sqrt{\Ncomp})$  when $\chimean \ll \chistd \sqrt{\Ncomp}$.
This means $\chimean$ leads to a linear correction of $(\barNunstable+1)/\Ncomp$.
\Figref{S-fig:linear_spinodal} shows that such a shift collapses the data for various $\chimean$.
Note that the shift $\chimean/(\pi \chistd \sqrt{\Ncomp})$ can be interpreted as the ratio between the governing parameter for $(\barNunstable+1)/\Ncomp$ in the identical interaction case, $\chimean/\Ncomp$, and that in the zero-mean random interaction case, $\chistd / \sqrt{\Ncomp}$.

We speculate that a similar linear relationship with respect to $\chimean$ holds for the phase count $\barNphase/\Ncomp$.
By comparing the governing parameters $\chimean - \ln(\Ncomp-1)$ and $\chistd \sqrt{\Ncomp}/\ln{\Ncomp}$, we conclude that the shift is proportional to $\chimean\ln(\Ncomp)/(\chistd\sqrt{\Ncomp})$.
We also include a fitting parameter $\lambda$ since we have little knowledge of the analytical form of $g_\mathrm{p}$, in contrast to $g_\mathrm{u}$.
\Figref{S-fig:linear_binodal}a shows that shifting the $\barNphase/\Ncomp$ data according to $\lambda\chimean\ln(\Ncomp)/(\chistd\sqrt{\Ncomp})$  collapses the data for various $\chimean$ over a broad range.
The fitting parameter is chosen to be $0.2$, independent of $\Ncomp$ (\Figref{S-fig:linear_binodal}b).
Note that these two linear relationships indicate that in the large $\Ncomp$ limit, the influences of mean interaction $\chimean$ is subtle, since both $1/(\pi \chistd \sqrt{\Ncomp})$ and $\ln{\Ncomp}/\sqrt{\Ncomp}$ are vanishingly small with large $\Ncomp$, consistent with \Figrefst{M-fig:all_std_and_mean} in the main text.

\subsection{Validation of the scaling laws at the center of the phase diagram}

\begin{figure}[t]
    \begin{center}
        \includegraphics[width = 0.8 \linewidth]{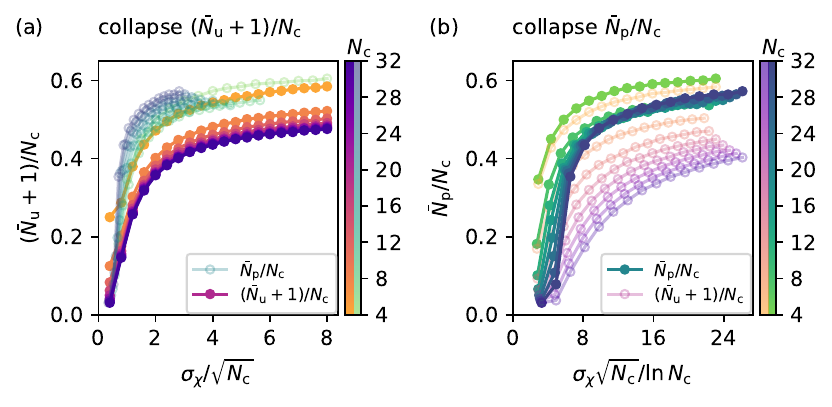}
    \end{center}
    \caption{
        \textbf{Average number of unstable modes and coexisting phases with random interactions but symmetric compositions.}
        Normalized average number of unstable modes, $(\barNunstable+1)/\Ncomp$ (orange-purple), and normalized average number of coexisting phases, $\barNphase/\Ncomp$ (green-blue), as a function of two different scalings of the standard deviation $\sigma_\chi$ of the interactions for various component counts $\Ncomp$ and $\chimean=0$.
        Each dot results from an average over $10^3$--$10^4$ pairs of random interaction matrices $\chi_{ij}$.
        The composition is fixed at $\barphi_i = \Ncomp $
    }
    \label{S-fig:binodal_sym_random}
\end{figure}

In the main text, we have averaged the phase count $\Nphase$ and the number $\Nunstable$ of unstable modes for random interactions uniformly over the entire phase diagram.
To test whether our derived scaling laws hold more generally, we also consider two other sampling methods:
First, we sample the boundary of the phase diagram uniformly, which effectively reduces the system to a system with $\Ncomp-1$ components, leaving all scaling laws unchanged when $\Ncomp \gg 1$.
Second, we only sample the center of the phase diagram at the symmetric compositions $\barphi_i = 1/\Ncomp$, where the scaling laws also still hold (\Figref{S-fig:binodal_sym_random}).
These scalings are also consistent with the earlier results for equal interactions given in Eqs.~(\ref{M-eqn:spinodal_symchi_symphi}) and (\ref{M-eqn:binodal_symchi_symphi_approx}).
Note that the scaling for $\barNunstable$ in this case is a direct result of random matrix theory~\cite{sear2003Instabilities}.
Taken together, these results suggest that the obtained scaling laws are not sensitive to the details of the composition sampling.
In particular, the observed discrepancy between $\Nphase$ and $\Nunstable$ reflects a general behavior of multicomponent mixtures.

\end{document}